\documentclass[12pt,useAMS, usenatbib]{mn2e}
\usepackage{amsmath}
\usepackage{amssymb}
\usepackage{graphicx}
\usepackage{subfigure}
\usepackage{mathrsfs}
\usepackage{bm}

\newcommand{\be}{\begin{equation}} 
\newcommand{\ee}{\end{equation}}

\newcommand{\ba}{\begin{eqnarray}} 
\newcommand{\ea}{\end{eqnarray}}

\newcommand{\D}{{\rm d}}

\newcommand{\svec}{\bm {\hat s}_\star}
\newcommand{\lvec}{\bm {\hat l}_1}
\newcommand{\jvec}{\bm {\hat j}}
\newcommand{\lpvec}{{\bm {\hat l}}_{\rm p}}
\newcommand{\ap}{a_{\mathrm{p}}}
\newcommand{\ep}{e_{\mathrm{p}}}
\newcommand{\aptilde}{\tilde{a}_{\mathrm{p}}}
\newcommand{\mper}{m_{\mathrm{p}}}

\def\go{\mathrel{\raise.3ex\hbox{$>$}\mkern-14mu
             \lower0.6ex\hbox{$\sim$}}}

\def\lo{\mathrel{\raise.3ex\hbox{$<$}\mkern-14mu
             \lower0.6ex\hbox{$\sim$}}}
\begin{document} 

\title[]{Teetering Stars: Resonant Excitation of Stellar Obliquities
  by Hot and Warm Jupiters with External Companions} 
\pagerange{\pageref{firstpage}--\pageref{lastpage}} \pubyear{2018}

\label{firstpage}

\author[K. R. Anderson \& D. Lai]{Kassandra R. Anderson\thanks{E-mail:
    kra46@cornell.edu} \& Dong Lai  \\ \\ Cornell Center for Astrophysics and Planetary Science, Department of Astronomy, Cornell University, Ithaca, NY 14853, USA}

\maketitle

\begin{abstract}
Stellar spin-orbit misalignments (obliquities) in hot Jupiter systems have been extensively probed.  Such obliquities may reveal clues about hot Jupiter dynamical histories.  Common explanations for generating obliquities include high-eccentricity migration and primordial disk misalignment.  This paper investigates another mechanism for producing stellar spin-orbit misalignments in systems hosting a close-in planet with an
external, modestly inclined companion.  Spin-orbit misalignment may be excited due to a secular resonance, occurring when the precession rate of the stellar spin axis (driven by the inner planet) becomes comparable to the nodal precession rate of the inner planet (driven by the companion). Due to the spin-down of the host star via magnetic braking, this resonance may be achieved during the star's main-sequence lifetime for a wide range of planet masses and orbital architectures.  Obliquity excitation is accompanied by a decrease in mutual inclination between the inner planet and perturber, and can thus erase high inclinations.  For hot Jupiters, the stellar spin axis is strongly coupled to the orbital axis, and obliquity excitation by a giant planet companion requires a strong perturber, usually located within 1-2 AU.  For warm Jupiters, the spin and orbital axes are more weakly coupled, and the resonance may be achieved for distant giant planet perturbers (at several to tens of AU).  Since warm Jupiters have a high occurrence rate of distant planetary companions with appropriate properties for resonant obliquity excitation, stellar obliquities in warm Jupiter systems may be common, particularly for warm Jupiters orbiting cool stars that have undergone significant spin-down. 
\end{abstract}


\section{Introduction}

Stellar spin-orbit misalignments (obliquities) in exoplanetary systems with a close-in planet have received significant attention in recent years.  The majority of detailed obliquity measurements have been conducted in hot Jupiter (HJ, giant planets with orbital periods less than ten days) systems via Rossiter-McLaughlin observations \citep[e.g.][]{winn2005,hebrard2008,narita2009,winn2009,triaud2010,albrecht2012}, yielding a wide range of sky-projected
obliquities, and even some retrograde systems \citep[][]{winn2015}.  Stellar obliquities provide a clue to the system's dynamical history, and may shed insight into planetary migration mechanisms.  Since there is still no consensus on how HJs
arrived at their short-period orbits, with several different proposed
migration theories, and even in-situ formation \cite[e.g.][]{dawson2018}, understanding the origins of stellar obliquities will further our understanding of HJ formation/migration.  In recent years, warm Jupiters (WJs, giant planets with
orbital periods between 10 and 300 days) have gained considerable
attention alongside HJs, and raise similar questions regarding their formation/migration.  Whether HJs and WJs arise
from a single or multiple formation channels is still an open question.

Low stellar obliquities are frequently attributed to either in-situ formation or disk-driven migration, in which the
planetary orbit shrinks due to gravitational torques from the protoplanetary disk.  In
contrast, high
obliquities may be attributed to high-eccentricity migration, in which gravitational interactions with other planets or a distant stellar companion raise the orbital eccentricity of a ``cold Jupiter'' to a large value, so that tidal dissipation at pericenter passages
leads to orbital decay \citep[e.g.][]{rasio1996,wu2003,fabrycky2007b,nagasawa2008,wu2011, beauge2012,naoz2012,petrovich2015a,petrovich2015b, anderson2016,munoz2016,hamers2017}.  High-eccentricity migration often results in large changes in orbital inclination, and even more extreme changes in the orientation of stellar spin axis itself \citep[][]{storch2014,storch2015,anderson2016,storch2017}, and is thus a natural way of producing large stellar spin-orbit misalignments.
However, as an alternative explanation for high obliquities, various works have investigated the possibility of
tilting the protoplanetary disk itself relative to the stellar spin axis. Such
primordial misalignments may allow for in-situ formation or disk-migration to result in
high obliquities, albeit with varying degrees of success \citep{bate2010,foucart2011,lai2011,batygin2012,batygin2013,lai2014,spalding2014,fielding2015,zanazzi2017}.
Given these results, exactly what obliquities inform us about planetary migration history
remains far from obvious.  

Thus far, primordial disk misalignment has been the main competitor to high-eccentricity migration in generating high stellar obliquities.  In this paper, we consider another mechanism in which a HJ/WJ that formed in-situ or through disk migration may attain in a high
stellar obliquity. This mechanism requires that the system host an
external, inclined planetary or stellar companion.  The companion induces nodal precession of the inner planet, causing its orbital axis to change direction; meanwhile, the oblate host star and
the inner planet (a HJ/WJ) experience a mutual torque, causing precession of both the stellar spin and orbital axes.  A secular resonance occurs when the spin axis precession frequency  (driven by the inner planet) is comparable to the orbital nodal precession frequency driven by the companion, potentially leading to large stellar obliquities, even for nearly aligned initial configurations \citep{lai2018}.  In this paper, we show that a system consisting of a host star, a HJ or WJ, and an outer companion may naturally pass through this secular resonance due to the spin-down of the star (by magnetic braking), and we examine to what extent large obliquities can be generated through this process.  We focus exclusively on planetary companions, but note that the results of this paper may also be applied to stellar companions.  In some scenarios, such companions may have previously induced high-eccentricity migration, leading to the formation of a HJ/WJ with a high obliquity.   However in this paper, we assume a formation process that resulted in a low initial obliquity (e.g. in-situ formation or migration within a protoplanetary disk aligned with the stellar equator), and identify the prospects for the companion to secularly raise the obliquity following the formation/migration.  The assumption of an initially low obliquity may be particularly appropriate for WJs, given that a high-eccentricity migration origin for such planets suffers from a number of difficulties (e.g. see \citealt{huang2016}, \citealt{antonini2016}, \citealt{AL2017}).

The role of external companions in affecting stellar obliquities has been studied before.  Some papers considered a spherical host star \citep[e.g.][]{kaib2011,becker2017} or a slowly rotating star \citep[e.g.][]{mardling2010}, so that the direction of the spin axis remains fixed or experiences little variation.  \cite{boue2014} and \cite{lai2018} examined the whole range of spin-orbit behaviors for oblate stars with a constant spin period, taking account of the spin axis changes due to 
gravitational torques from the inner planet.  Our paper builds upon these works by exploring the dynamical evolution on Gyr timescales, so that the stellar spin-down (by magnetic braking) plays an important role in the evolution of the stellar obliquity.  In addition, we show that the resulting spin-orbit dynamics and obliquity excitation act to decrease the mutual inclination between the two planets.

The resonant obliquity excitation mechanism studied in this paper requires that the HJ/WJ have external companions with certain ranges of masses and orbital separations, as well as modest inclinations.  Distant planetary companions to HJs and WJs are common, with estimated occurrence rates $50
\%$ for WJs and up to $80 \%$ for HJs \citep{bryan2016}.  There is
a growing number of systems with well-characterized orbits for the
companion, especially for WJs \citep[see][for a recent compilation of WJs with external companions]{antonini2016}.  Mutual inclinations in giant planet systems remain far less constrained, although recent observations
are beginning to probe individual systems, with several in high-inclination configurations \citep{mills2017, masuda2017}.  Upcoming {\it Gaia} results may provide further constraints on mutual inclinations of giant planet systems \citep[e.g.][]{perryman2014}. 

 We note that the present sample of stellar obliquity
measurements is limited mostly to HJs. The results of this
paper show that high obliquities may be common for
WJs with external companions, regardless of their formation history.

This paper is organized as follows.  In Section \ref{sec:setup} we outline the problem setup and review the relevant spin-orbit dynamics.  In Section \ref{sec:ideal} we explore in detail the process of resonant obliquity excitation using a somewhat idealized model, where the stellar spin angular momentum is much less than the inner planet orbital angular momentum.  Such a model serves as a starting point in understanding the dynamics of more realistic systems, with comparable spin and orbital angular momenta.  In Section \ref{sec:numerical} we relax the assumption of small spin angular momentum, and numerically explore the parameter space for HJs/WJs with various types of external companions.  We summarize and conclude in Section \ref{sec:conclusion}.

\section{Setup \& Classification of Dynamical Behavior}
\label{sec:setup}
We consider an oblate star of mass $M_\star$, radius $R_\star$, and spin
period $P_\star$, hosting a close-in giant planet $m_1$ (either a HJ or WJ) in a circular orbit with
semi-major axis $a_1$, and a distant 
perturber $\mper$, with semi-major axis $\ap$, eccentricity $e_{\rm p}$, and inclination $I$ relative to
the orbit of the inner planet.  Both planets are considered as point masses. The star has spin angular momentum $S_\star$, and the inner planet and the perturber have orbital angular momenta $L_1$ and $L_{\rm p}$ respectively. The quadrupole-order secular equations of
motion for the spin unit vector $\svec = {\bf S}_\star/S_\star$ and the orbital angular momentum unit vectors
$\lvec = {\bf L}_1/L_1$ and $\lpvec = {\bf L}_{\rm p}/L_{\rm p}$ are
\ba
\frac{\D \svec}{\D t} & = & \omega_{\star 1} (\svec \cdot \lvec) (\svec \times \lvec)  \label{eq:dsdt} \\ 
\frac{\D \lvec}{\D t} & = &  \omega_{{1 \rm p}} (\lvec \cdot \lpvec) (\lvec
\times \lpvec) + \frac{S_\star}{L_1} \omega_{\star 1} (\lvec \cdot \svec) (\lvec \times \svec)  \label{eq:dldt}   \\
\frac{\D \lpvec}{\D t} & = & \frac{L_1}{L_{\rm p}} \omega_{1 {\rm p} } (\lpvec \cdot \lvec) (\lpvec
\times \lvec) \label{eq:dlpdt} ,  
\ea
where the relevant precession frequencies are
\be
\omega_{\star 1} = \frac{3 k_{q \star}}{2 k_\star} \bigg(\frac{m_1}{M_\star}
\bigg) \bigg(\frac{R_\star}{a_1} \bigg)^3 \Omega_\star,
\label{eq:starprecess}
\ee
and 
\be
\omega_{1 {\rm p}} = \frac{3 \mper}{4 M_\star} \bigg(\frac{a_1}{\aptilde} \bigg)^3
n.
\label{eq:orbitprecess}
\ee
In equation (\ref{eq:starprecess}), 
$\Omega_\star = 2 \pi /P_\star$ is the angular frequency of the star,
and $k_{\star}$ and $k_{q \star}$ are related to the stellar moment of
inertia and quadrupole moment (see \citealt{lai2018}) for precise
definitions.  In equation (\ref{eq:orbitprecess}), $n = \sqrt{G
  M_\star/a_1^3}$ is the orbital mean motion of the inner planet, and we have defined an effective semi-major axis of the perturber\footnote{The perturber properties enter mainly in the combination $\aptilde / \mper^{1/3}$; however, we note that additional dependence is introduced through the ratio $L_1 / L_{\rm p}$.},
  \be
  \aptilde \equiv \ap \sqrt{1 - e_{\rm p}^2}.
  \label{eq:aptilde}
  \ee
Note that in equations (\ref{eq:dsdt}) - (\ref{eq:dlpdt}) we have neglected the coupling between the star and outer planet \citep[see][]{lai2018}, which induces precession of $\svec$ at a rate $\omega_{\star {\rm p}} \sim (\mper / M_\star)(R_\star^3/\aptilde^3) \Omega_\star$, and is completely negligible for this problem.

The dynamical behavior of the system can be described as follows: $\svec$ and $\lvec$ mutually precess around the axis defined by ${\bf S}_\star + {\bf L}_1$, while, $\lvec$ and $\lpvec$ undergo mutual
precession around the total orbital angular momentum axis defined by ${\bf L}_1 +
{\bf L}_{\rm p}$. The evolution of $\svec$ due to the forcing of $\lvec$ (which is
itself being forced by $\lpvec$) is therefore complicated, depending crucially
on the relative precession rates $\omega_{\star 1}$ and $\omega_{1 {\rm p}}$,
as well as the angular momentum ratio $S_\star/L_1$.  For a rapidly rotating
star, $S_\star$ can be comparable to $L_1$, so that the back-reaction torque
from the oblate star on the orbit is non-negligible.

Meanwhile, the rotation rate of the star $\Omega_\star$ decreases due to magnetic braking. We adopt the Skumanich law ($\dot{\Omega}_\star \propto - \Omega_\star^3$; see \citealt{skumanich1972}, \citealt{bouvier2013}) for the stellar spin-down,
so that the spin frequency as a function of time is given by
\be
\Omega_\star = \frac{\Omega_{\star,0}}{\sqrt{1 + \alpha_{\rm MB}
    \Omega_{\star,0}^2 t}},
\label{eq:magbraking}
\ee
where $\Omega_{\star,0}$ is the initial spin rate and $\alpha_{\rm MB}$ is a
constant, calibrated such that the rotation period reaches $\sim 30$
days at an age $\sim 5$ Gyr.  In this work we adopt $\alpha_{\rm MB} = 1.5 \times 10^{-14}$
yr, appropriate for solar-mass stars \citep{barker2009}. 

The qualitative spin-orbit dynamics depend on the relevant precession rates \citep{boue2014,lai2018}.  \cite{lai2018} describe the
spin-orbit dynamics by introducing the
dimensionless parameter $\epsilon_{\star 1}$, which, for a giant
planet on a short period orbit can be approximated as
\ba
\epsilon_{\star 1} & = & \frac{\omega_{1 {\rm p}} - \omega_{\star {\rm p}} }{\omega_{\star 1} (1 + S_\star/L_1)} \simeq  \frac{\omega_{1 {\rm p}}}{\omega_{\star 1}}
\bigg(\frac{1}{1 + S_\star/L_1} \bigg) \nonumber \\
& \simeq & 1.25 \left(\frac{k_\star}{6 k_{q \star}} \right) \left(\frac{\mper}{m_1}\right) \left(\frac{a_1}{0.04 \
    {\rm AU}}\right)^{9/2} \left(\frac{\aptilde}{1 \ {\rm AU}} \right)^{-3}
 \nonumber \\ 
& & \times \left( \frac{P_\star}{30 \ {\rm d}} \right) \bigg(\frac{M_\star}{M_\odot} \bigg)^{1/2} \bigg(\frac{R_\star}{R_\odot} \bigg)^{-3}  \left(\frac{1}{1 +
  S_\star / L_1} \right).
\label{eq:eps}
\ea
We summarize the key points from \cite{lai2018} here:
(i) If $\epsilon_{\star 1} \ll 1$, $\svec$ and $\lvec$ are strongly
coupled and undergo rapid
mutual precession, and the spin-orbit angle $\theta = \theta_{\star 1}$ (the
angle between $\svec$ and $\lvec$) satisfies $\theta
\simeq$ constant.  If $\svec$ and $\lvec$ are initially aligned, spin-orbit misalignment cannot be generated when $\epsilon_{\star 1} \ll 1$.
(ii) If $\epsilon_{\star 1} \gg 1$, $\svec$ and $\lvec$ are weakly
coupled, and both precess around the total orbital angular momentum
axis, but $\lvec$ precesses at a much faster rate than $\svec$.  As a
result, the spin-orbit angle varies between a minimum and a maximum
value.  For an initially aligned system, and when $L_1 \ll
L_{\rm p}$, the spin-orbit angle varies in the range $0 \lesssim
\theta  \lesssim 2 I$ over a precession cycle. (iii)
If $\epsilon_{\star 1} \simeq 1$, a secular
spin-orbit resonance occurs due to the commensurability between the precession frequencies $\omega_{\star 1}$ and $\omega_{1 {\rm p}}$, and $\theta$ may grow to a
large value.  See also \cite{lai2017} (particularly Appendix A) and \cite{pu2018} (Section 2.2) for more theory on the details of this resonance.

\cite{lai2018} considered systems where the stellar spin-rate was
held constant.  In this case, resonant excitation of obliquity requires an outer perturber with
somewhat fine-tuned properties, due to the strong dependence of
$\epsilon_{\star 1}$ on $\aptilde$.  However, over Gyr timescales, the
stellar spin period is reduced by a factor of $\sim 10$ due to
magnetic braking, so that $\epsilon_{\star 1}$ is a function of
time.  Systems that begin with spin-orbit alignment and
$\epsilon_{\star 1} \ll 1$ (in the strong-coupling regime) may eventually cross $\epsilon_{\star 1} \simeq 1$ due to
magnetic braking, so that
$\theta$ resonantly grows.  After the resonance is
encountered, the system enters the weak-coupling regime, with
$\theta$ varying between a minimum and a maximum.  We will
show in this paper that the ``final'' range of variation of the spin-orbit angle
(following resonant excitation)
depends on the spin history of the system.

\begin{figure}
\centering 
\includegraphics[scale=0.5]{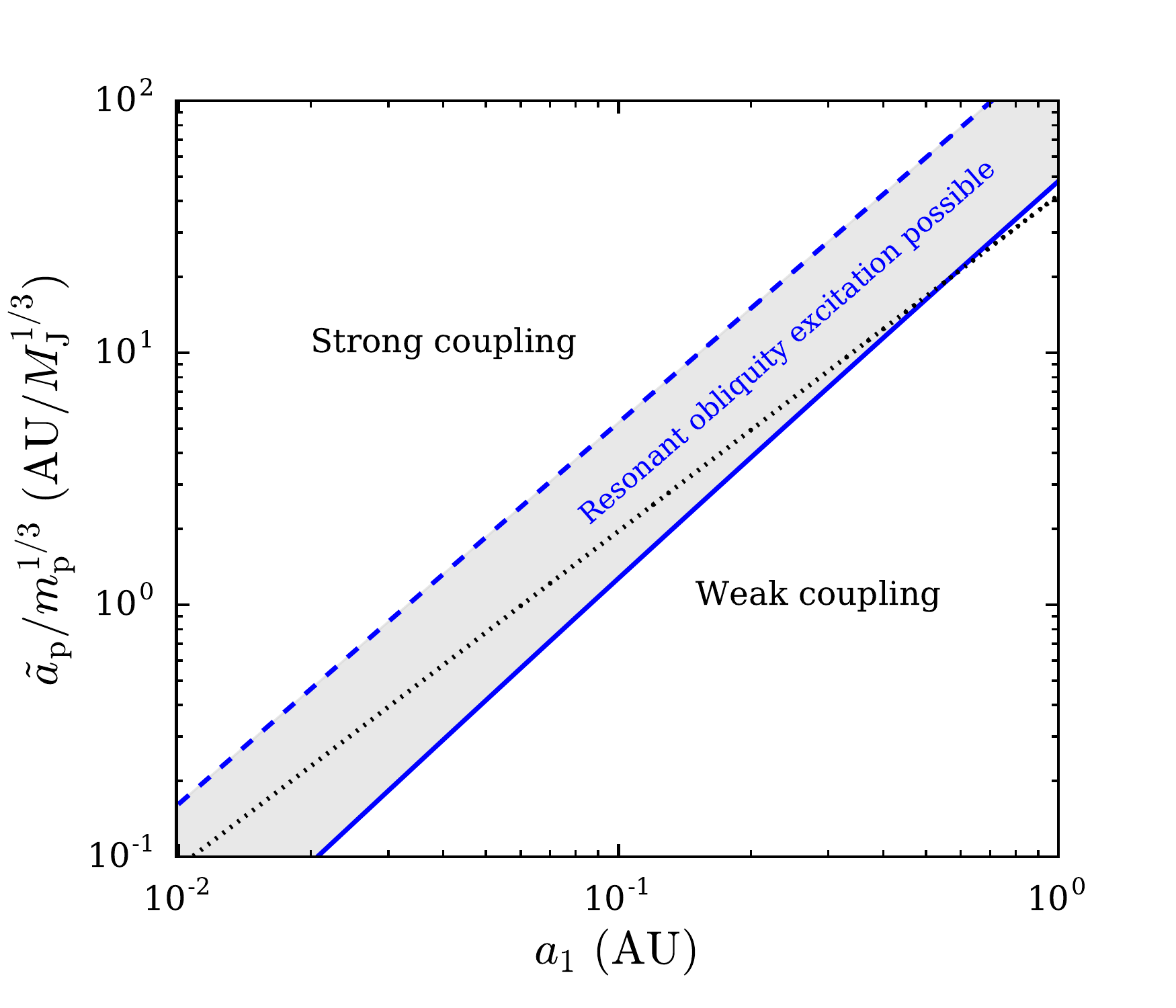}
\caption{Parameter space for resonant excitation of the stellar obliquity to
  be possible (grey region) for an inner planet with mass $m_1 = M_{\rm J}$,
  in terms of its semi-major axis $a_1$, and the perturber ``strength'' $\aptilde/\mper^{1/3}$, where $\aptilde$ is given by equation (\ref{eq:aptilde}).  The blue solid (dashed) lines
  indicate constant $\epsilon_{\star 1} = 1$, with $P_\star =
  1$ (30) days.  The black dotted line indicates the maximum value of $\aptilde / \mper^{1/3}$ for Lidov-Kozai
  eccentricity oscillations to be possible, given a sufficiently high
  inclination.  }
\label{fig:res_cond}
\end{figure}

Resonant excitation of stellar obliquities requires that the system
initially satisfy $\epsilon_{\star 1} \lesssim 1$. In addition, in order for
the resonance to be encountered within a reasonable time (within, say
$5$ Gyr), we require that $\epsilon_{\star 1} (t = 5 {\rm Gyr})
\gtrsim 1$.  For an inner planet with $m_1 = M_{\rm J}$ and various values of $a_1$, and an initial stellar spin period of
$1$ day (roughly the lower limit obtained from observations of T-Tauri stars), the range of perturber ``strength'' ($\aptilde/\mper^{1/3}$) allowing
resonant obliquity excitation may be identified, shown as the shaded grey region in Fig.~\ref{fig:res_cond}.  Note that this region allowing resonant growth narrows slightly with increasing initial stellar spin period (the lower boundary, solid blue line).  The upper boundary (dashed blue line) is independent of initial spin period, because solar-type stars ``forget'' their initial spin periods after several hundred Myr.  The boundaries of parameter space allowing resonant obliquity excitation in Fig.~\ref{fig:res_cond} are
approximate.  In Section 4 we perform a
thorough numerical exploration of the parameter space and numerically confirm that the shaded region in Fig.~\ref{fig:res_cond} does indeed identify the parameter space available for resonant obliquity growth.

Since
a sufficiently inclined perturber can also excite the inner planet eccentricity,
we plot the necessary condition for quadrupole-level Lidov-Kozai eccentricity oscillations in Fig.~\ref{fig:res_cond}.  This arises from requiring that the rate of apsidal precession due to general relativity is sufficiently slow compared to that induced by the perturber \citep[see, e.g. equation 29 of][]{anderson2017}.  For a given value of $a_1$, perturbers below the black dotted line may induce Lidov-Kozai oscillations. Note that this condition is necessary for Lidov-Kozai cycles to develop, but not sufficient, because a minimum mutual inclination ($I_{\rm LK,min}$) is also required.\footnote{An upper boundary $I_{\rm LK,max}$ also exists, so that Lidov-Kozai cycles also require $I_0 < I_{\rm LK,max}$ (with $I_{\rm LK,max}$ retrograde).  However, this upper boundary is probably irrelevant for planetary companions, as such retrograde inclinations are not easily produced.} In the idealized scenario where apsidal precession from general relativity and other ``short-range-forces'' are neglected, $I_{\rm LK,min} \simeq 40^\circ$.  Inclusion of short-range forces causes $I_{\rm LK,min}$ to exceed $40^\circ$, often by a considerable amount. 
Inspection of Fig.~\ref{fig:res_cond} reveals that there is some
parameter space for HJs that may allow both resonant obliquity excitation and Lidov-Kozai
cycles (although note that the required perturber must be quite close/strong, and often in conflict with observational constrains of companions to HJs; \citealt{huang2016}).  Since this paper assumes circular orbits for HJs/WJs, for simplicity we will always restrict out attention to initial inclinations less than $40^\circ$ to ensure that Lidov-Kozai oscillations do not arise.  However, we note that the qualitative spin-orbit dynamics discussed in this paper will often hold for higher inclinations, as long as $I_0 < I_{\rm LK,min}$.  If Lidov-Kozai oscillations do occur, then the evolution of the stellar spin axis becomes chaotic \citep{storch2014,storch2015}, which may allow the full range of obliquities ($0^\circ - 180^\circ$) to be explored.  This issue is beyond the scope of this paper.

In this paper we neglect tides raised by the planet on the host star, which may lead to orbital decay and damping of obliquities.  The timescale for tides to reduce the semi-major axis of a planet around a solar-type star is
\be
t_a \simeq 1.3 \times 10^{11} {\rm yr} \bigg(\frac{Q_{\star}^{'}}{10^7} \bigg) \bigg(\frac{m_1}{M_J} \bigg)^{-1} \bigg(\frac{a_1}{0.04 \ {\rm AU}} \bigg)^{13/2},
\label{eq:tides}
\ee
where $Q_{\star}^{'} = 3Q_\star/(2 k_2)$ is the reduced tidal quality factor.  The timescale for obliquity decay is $t_{\theta} \sim (S_\star/L_1)t_a$ \citep[see][for a discussion on the relation between $t_a$ and $t_\theta$]{lai2012}.  Although stellar tides can be important for HJs that are massive and/or in sufficiently short-period orbits, tides are unimportant for HJs that may experience changes in stellar obliquity from external companions.  For example, using the canonical values of $Q_{\star}^{'}$ and $m_1$ in equation (\ref{eq:tides}),  a HJ with $a_1 \simeq 0.02$  AU has a tidal decay timescale $t_a \simeq 1.4$ Gyr, so that tides may indeed sculpt the semi-major axis and stellar obliquity over the $\sim $ Gyr timescales of interest in this paper.  However, such a system will always be in the strong-coupling regime ($\epsilon_{\star 1} \ll 1$) throughout the main-sequence lifetime of the star, unless the system hosts an extremely strong external perturber, with $\aptilde / \mper^{1/3} \lesssim 0.4 {\rm AU}/M_J^{1/3}$.  Since HJs have been shown to lack such companions \citep{huang2016}, we expect tides (for $Q_{\star}^{'} \sim 10^6 - 10^7$) to be completely negligible for the systems of interest in this paper.  

The above estimate of the tidal timescales (with $Q_{\star}^{'} \sim 10^7$) assumes equilibrium tides \citep{zahn1977}, where the source of dissipation is damping by turbulent viscosity in the convective region.  \cite{mathis2015} and \cite{bolmont2016} have recently suggested that $Q_{\star}^{'}$ may briefly attain a much lower value ($Q_{\star}^{'} \sim 10^{3.5}$) for rapidly rotating pre-main-sequence stars, due to excitation and damping of inertial waves in the convective shell (see \citealt{ogilvie2013} for the original calculation based on idealized two-zone stellar models).  The ensuing orbital evolution of HJs in this scenario was recently explored by \cite{heller2018}. While intriguing, such a tidal treatment is beyond the scope of this present paper.

\section{Spin-Orbit Dynamics when ${\bf S_\star \ll L_1}$}
\label{sec:ideal}
In this section we review and develop some analytic results in order
to gain insight into the spin-orbit dynamics.  We
consider a limiting case where $S_\star/L_1 \ll 1$, and defer the discussion of comparable $S_\star$ and $L_1$ to Section \ref{sec:compSL}.  Since realistic HJs and WJs often satisfy $S_\star \sim L_1$, especially when the host star is young, the
following discussions are somewhat idealized, but shed insight into the dynamical evolution of more complicated systems.  Readers interested in the quantitative results and conclusions for more typical HJ and WJ
systems (with $S_\star \sim L_1$) are referred to Sections 4 and 5.    

The spin-axis dynamics in the limit $S_\star/L_1 \ll 1$ has been studied in a variety of
contexts, and is related to the well-known Cassini
state problem \citep[e.g.][]{colombo1966,peale1969,
  peale1974,ward1979,henrard1987,ward2004,fabrycky2007}.  In the following we
review the relevant spin-axis dynamics and Cassini state theory.

\subsection{Cassini States \& Phase Space Structure}

When $S_\star \ll L_1$, the back-reaction torque of the spin on the
orbit vanishes, so that the orbital axis $\lvec$ is unaffected by $\svec$, 
and simply undergoes nodal precession due to $\mper$.  The invariable plane is thus defined by the unit vector $\jvec$, in the direction of the total orbital angular momentum ${\bf J} = {\bf L}_1 + {\bf L}_{\rm p}$, and $\lvec$ precesses around $\jvec$, with constant inclination $I'$ according to
\be
\frac{\D \lvec}{\D t} = g (\jvec \times \lvec),
\ee
where the precession frequency $g$ is
\be
g = -\frac{J}{L_{\rm p}} \omega_{1 {\rm p}} \cos I,
\ee
and where $I$ is the angle between $\lvec$ and $\lpvec$.  For ease of notation, we will work in the limit $L_1 \ll L_{\rm p}$ for the remainder of Section 3, so that $I' \to I$ and $g \to -\omega_{1 \rm p} \cos I$, but the following results are valid for comparable $L_1$ and $L_{\rm p}$, with $\jvec$ replacing $\lpvec$ and $I'$ replacing $I$.

Following standard procedures, we transform to the frame rotating with
frequency $g$, where $\lvec$ is fixed and directed along the
$z$-axis.  In this rotating frame, $\lpvec$ is fixed, and $\svec$ evolves according to
\be
\left( \frac{\D \svec}{\D t} \right)_{\rm rot} = \alpha (\svec \cdot \lvec) (\svec \times \lvec) + g (\svec \times \lpvec).
\label{eq:sdot}
\ee
In equation (\ref{eq:sdot}), we have adopted standard notation, where the
spin precession constant $\alpha = \omega_{\star 1}$. We may rescale time such that $\tau = \alpha t$; thus the spin dynamics only
depend on the ratio $g / \alpha$ and $I$. 

\begin{figure}
\centering 
\includegraphics[scale=0.5]{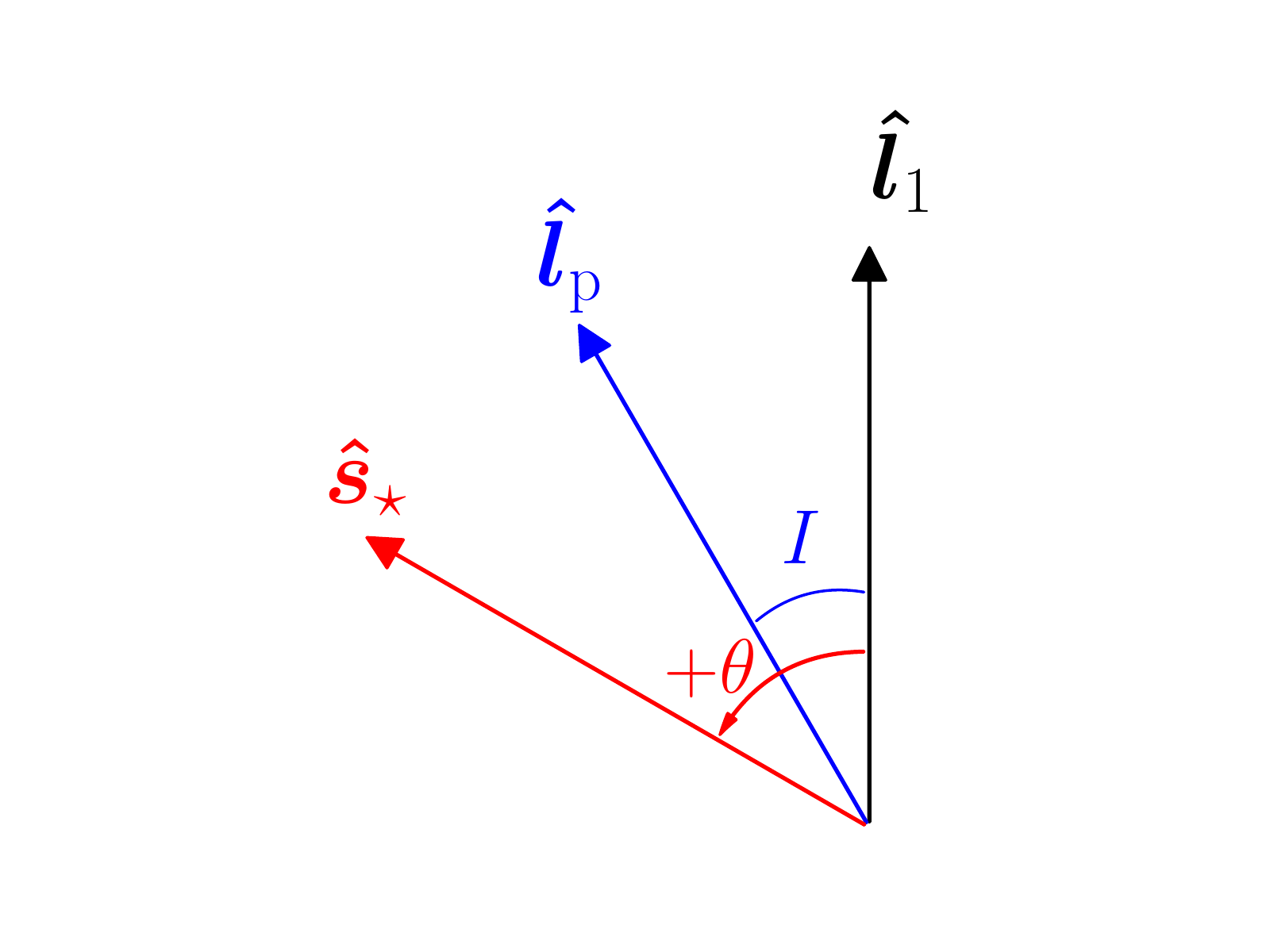}
\caption{Cartoon illustration of the Cassini state configuration (with
  $\svec$, $\lvec$, and $\lpvec$ located in the same plane) and the adopted
  sign convention for $\theta$.
When $\theta > 0$, $\svec$ and $\lpvec$ are located on the same side of $\lvec$
(as shown, corresponding to $\theta_2$).
When $\theta < 0$, $\svec$ and $\lpvec$ are on opposite sides of $\lvec$ (corresponding to $\theta_{1,3,4}$).}
\label{fig:vectors}
\end{figure}

\begin{figure*}
\centering 
\includegraphics[width=\textwidth]{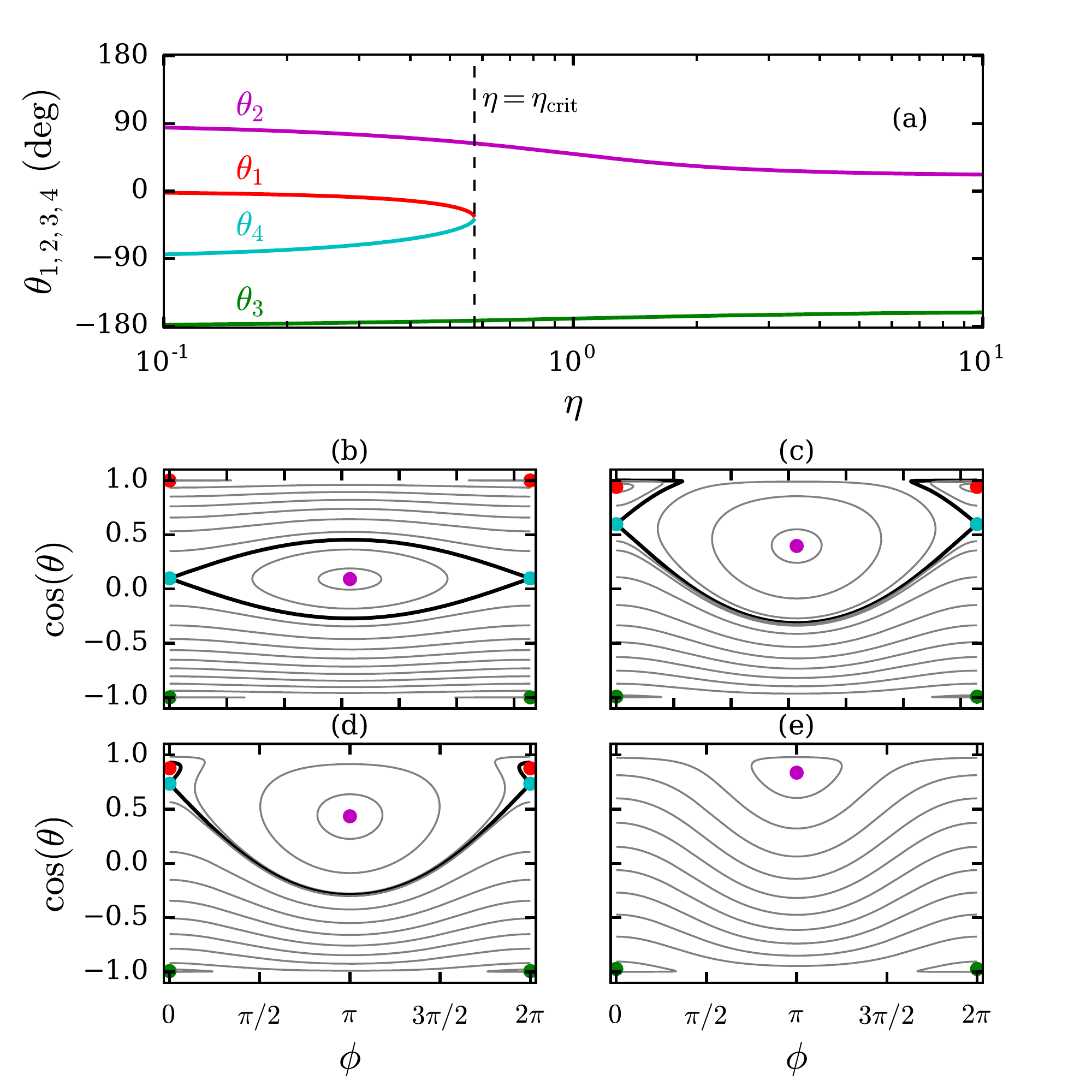}
\caption{{\it Panel (a)}: Cassini states $\theta_{1,2,3,4}$ versus $\eta =
  |g|/\alpha$, with fixed $I = 20^\circ$. {\it Panels (b) - (e)}:
 Phase-space portraits for various values
  of $\eta$.  {\it Panel (b)}: $\eta = 0.1$.  For such a small 
  $\eta$, the separatrix (thick black curve passing through
  $\theta_4$) is narrow, and almost all trajectories outside of the
  separatrix circulate, except for small librations very close to $\theta_1$ and
  $\theta_3$. {\it Panel (c)}: $\eta = 0.5$.  As $\eta$ increases, the
  separatrix expands until it touches $\cos \theta = 1$ (not shown here), after which
  the shape of the separatrix abruptly changes, now enclosing
  $\theta_1$. {\it Panel (d)}:  $\eta
  = 0.561$.  Phase space just before $\theta_1$
  and $\theta_4$ merge.  {\it Panel (e)}: $\eta =  2$. Phase space after $\theta_1$
  and $\theta_4$ have merged, so that the only remaining Cassini states are $\theta_2$ and $\theta_3$.}
\label{fig:phase_space}
\end{figure*}

The dynamical evolution of $\svec$ can be specified by the obliquity $\theta$ (the angle between $\svec$ and $\lvec$), and the phase angle $\phi$ (the longitude of ascending node of the stellar equator in the rotating frame).  Note that $\cos \theta$ and $\phi$ are canonical variables
for the Hamiltonian governing this system, which is given by
\be
\mathcal{H} =  -\frac{\alpha}{2} (\svec \cdot \lvec)^2  - g (\svec \cdot \lpvec). 
\label{eq:ham}
\ee
The equilibrium, or Cassini states, are obtained by setting equation (\ref{eq:sdot}) to zero.  This requires that $\svec$, $\lvec$, and $\lpvec$ are coplanar (with $\phi = 0$ or $\pi$), implying that $\svec$ and $\lvec$ precess at the same rate in inertial space around $\lpvec$.  The Cassini state obliquities satisfy
\be
\frac{g}{\alpha} \sin (\theta - I) + \sin \theta \cos \theta = 0.
\label{eq:cass}
\ee
Equation (\ref{eq:cass}) has either
two or four solutions, depending on the values of $g/\alpha$ and $I$.  Following
standard convention and nomenclature (See Figs.~\ref{fig:vectors} and \ref{fig:phase_space}), Cassini states 1, 3 and 4 ($\theta_{1,3,4} < 0$) occur when $\svec$ and
$\lpvec$ are on opposite sides of $\lvec$ ($\phi = 0$), while Cassini
state 2 ($\theta_2 > 0$) occurs when $\svec$ and $\lpvec$ are on the same side of
$\lvec$ ($\phi = \pi$).    

For convenience, we define $\eta \equiv |g| / \alpha$.  Note that
$\eta$ is related to the parameter $\epsilon_{\star 1}$
introduced in Section 2 (see equation \ref{eq:eps}), by $\eta = \epsilon_{\star 1} |\cos I|$ (for $S_\star / L_1 \ll 1$).  Thus, $\eta \ll$
corresponds to strong coupling between $\svec$ and $\lvec$, while $\eta \gg 1$ corresponds to weak
coupling.  When $\eta < \eta_{\rm crit}$, with
\be
\eta_{\rm crit} = (\sin^{2/3} I + \cos^{2/3} I )^{-3/2},
\ee
all four Cassini states exist, whereas when $\eta > \eta_{\rm crit}$, only $\theta_2$ and $\theta_3$ exist (see Fig.~\ref{fig:phase_space}).

The Cassini states $\theta_1, \theta_2, \theta_3$ are stable, while $\theta_4$ is unstable and
lies along a separatrix in the underlying phase space ($\cos \theta,\phi$).  In Fig.~(\ref{fig:phase_space}a) the Cassini states are depicted as a
function of $\eta$ with fixed $I = 20^\circ$.   When $\eta = \eta_{\rm crit}$, $\theta_1$ and $\theta_4$ merge and destroy each other.

The phase space structure (contours of constant $\mathcal{H}$, see equation [\ref{eq:ham}]) is shown in Fig.~(\ref{fig:phase_space}b - \ref{fig:phase_space}e) for increasing values of $\eta$.  For
values of $\eta \ll 1$ (Fig.~\ref{fig:phase_space}b), the separatrix (which passes through $\theta_4$ and
encloses $\theta_2$)
is relatively narrow, and most of the trajectories circulate (over $\phi$)
with little variation of $\cos \theta$, although librating
trajectories exist close to $\theta_1$, $\theta_2$ and $\theta_3$.  As $\eta$ increases, the
separatrix widens, and eventually when $\eta$ is close to, but less
than $\eta_{\rm crit}$, the ``top'' of the separatrix touches $\cos
\theta = 1$.  As $\eta$ increases further, the shape of the separatrix changes, and encloses $\theta_1$
(see Fig.~\ref{fig:phase_space}c).  The phase space just before $\theta_1$
and $\theta_4$ merge is shown in Fig.~\ref{fig:phase_space}d, and just after merging in Fig.~\ref{fig:phase_space}e.  When $\eta > \eta_{\rm crit}$, the only prograde Cassini state is $\theta_2$.

\subsection{Spin-Orbit Resonance and Separatrix Crossing}
We next consider the scenario where $\eta$ slowly increases with
time.  The exact form of $\eta (t)$ is
unimportant, as long as $\eta$ increases slowly compared to all the
precession timescales.

When $\eta$ changes slowly, the area of the phase-space trajectory
is constant, so that
\be
A \equiv \oint \cos \theta \,d \phi = {\rm constant}.
\label{eq:A}
\ee
Equation (\ref{eq:A}) only holds as long as there are no
abrupt changes in the phase space structure (e.g. if $\eta$ crosses
$\eta_{\rm crit}$, $A$ is not conserved).

\begin{figure*}
\centering 
\includegraphics[width=\textwidth]{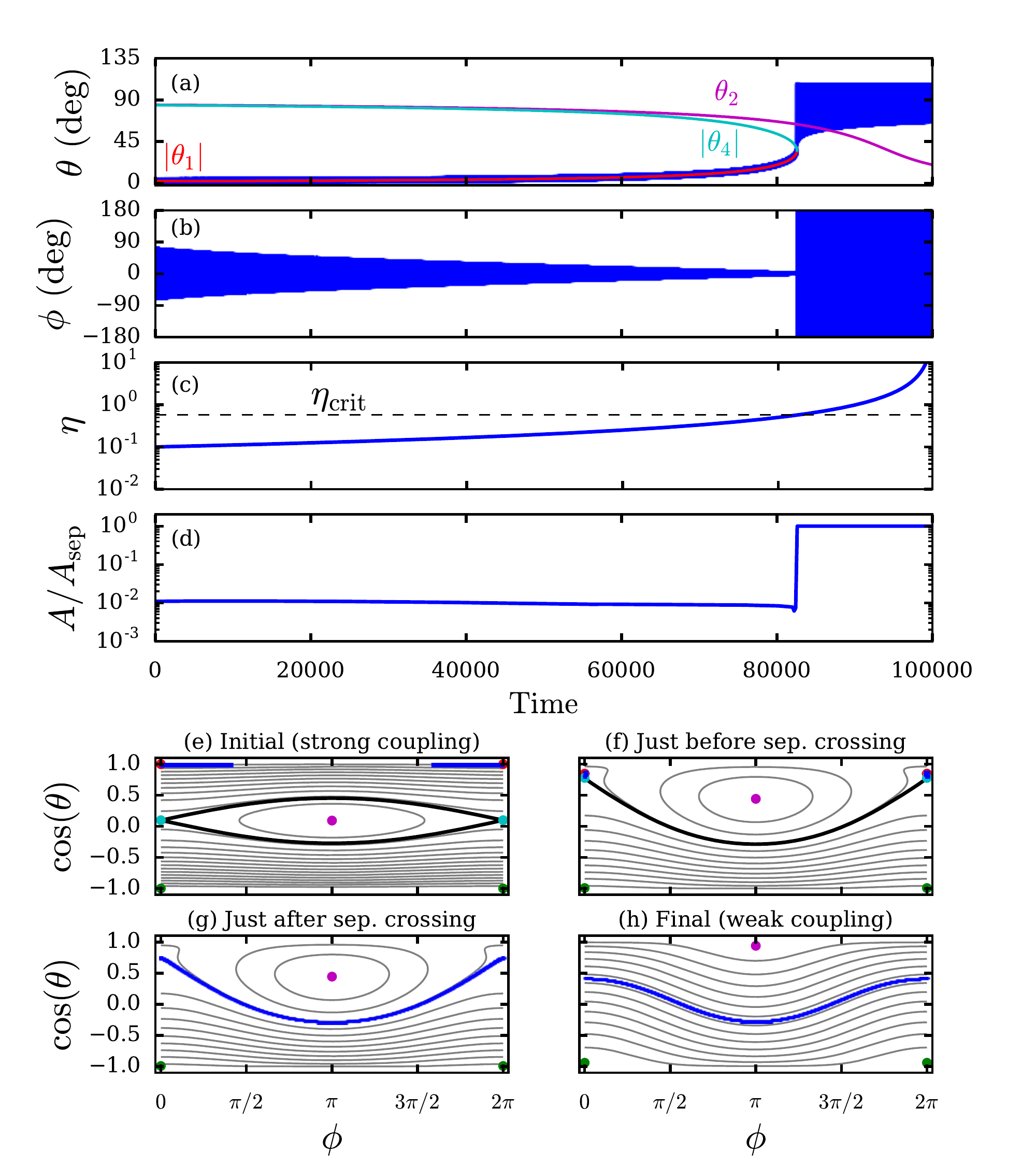}
\caption{Spin evolution with slowly increasing $\eta$, for a system with $S_\star/L_1 \ll 1$, so that $I = $ constant (as discussed in Section 3).  All blue curves show the result of a numerical integration of equation (\ref{eq:sdot}), where $\eta = |g|/ \alpha$ slowly increases with time according to $\eta \propto (1 - c t)^{-1}$, where $c$ is a constant. Panels (a)-(d) show the obliquity ($\theta$), the precessional phase of $\svec$ in the rotating frame ($\phi$), the coupling parameter $\eta$, and phase space area $A$ (normalized by the area of separatrix when $\eta = \eta_{\rm crit}$).  Panels (e)-(h) show the phase space
  trajectory obtained from the numerical integration (blue curves),
  along with the underlying phase space (grey contours), fixed points, and separatrix (thick black curve).  Panel (e): Initial phase space, when $\eta = 0.1$ and the spin axis librates around $\theta_1$.  Panel (f): Phase space just before $\theta_1$ and $\theta_4$ merge, with the spin axis tightly enclosed by the separatrix and librating around $\theta_1$ (compare with Fig.~\ref{fig:phase_space}c).  Panel (g): Phase space just after $\theta_1$ and $\theta_4$ merge.  The spin axis now circulates around the only remaining prograde Cassini state, $\theta_2$.  Panel (h): Phase space when $\eta \gg 1$, showing the final degree of obliquity variation, which varies in the range $2 I$. }
\label{fig:time_ev_phase_space}
\end{figure*}

A numerical integration of equation (\ref{eq:sdot}) with slowly increasing $\eta$ is
shown in Fig.~\ref{fig:time_ev_phase_space}, where initially $\theta \simeq 0$ and $\eta \ll 1$.  At early times, the spin axis is strongly coupled to
the orbital axis, so that $\theta$ remains nearly constant, and the
system librates around Cassini state $1$ ($\theta_1$), and the
area of the trajectory ($A$) is small.  As $\eta$
increases, $\theta_1$ increases in magnitude, and
the spin axis continues to librate around $\theta_1$ while preserving
phase-space area.  Eventually, when $\eta = \eta_{\rm crit}$, $\theta_1$ merges
with $\theta_4$, and the system is forced to cross the separatrix.  At the separatrix crossing, the obliquity undergoes a rapid increase and the phase-space area increases by a
factor of $\sim 100$.  After the separatrix crossing, $\phi$
circulates, and $\theta$ varies between a maximum and minimum value,
determined by the area of the separatrix when $\eta = \eta_{\rm crit}$.
The system continues to evolve, while preserving the new, much larger
phase space area. We refer to the process of rapid obliquity growth during the separatrix crossing as resonant excitation of the obliquity.

\begin{figure}
\centering 
\includegraphics[scale=0.45]{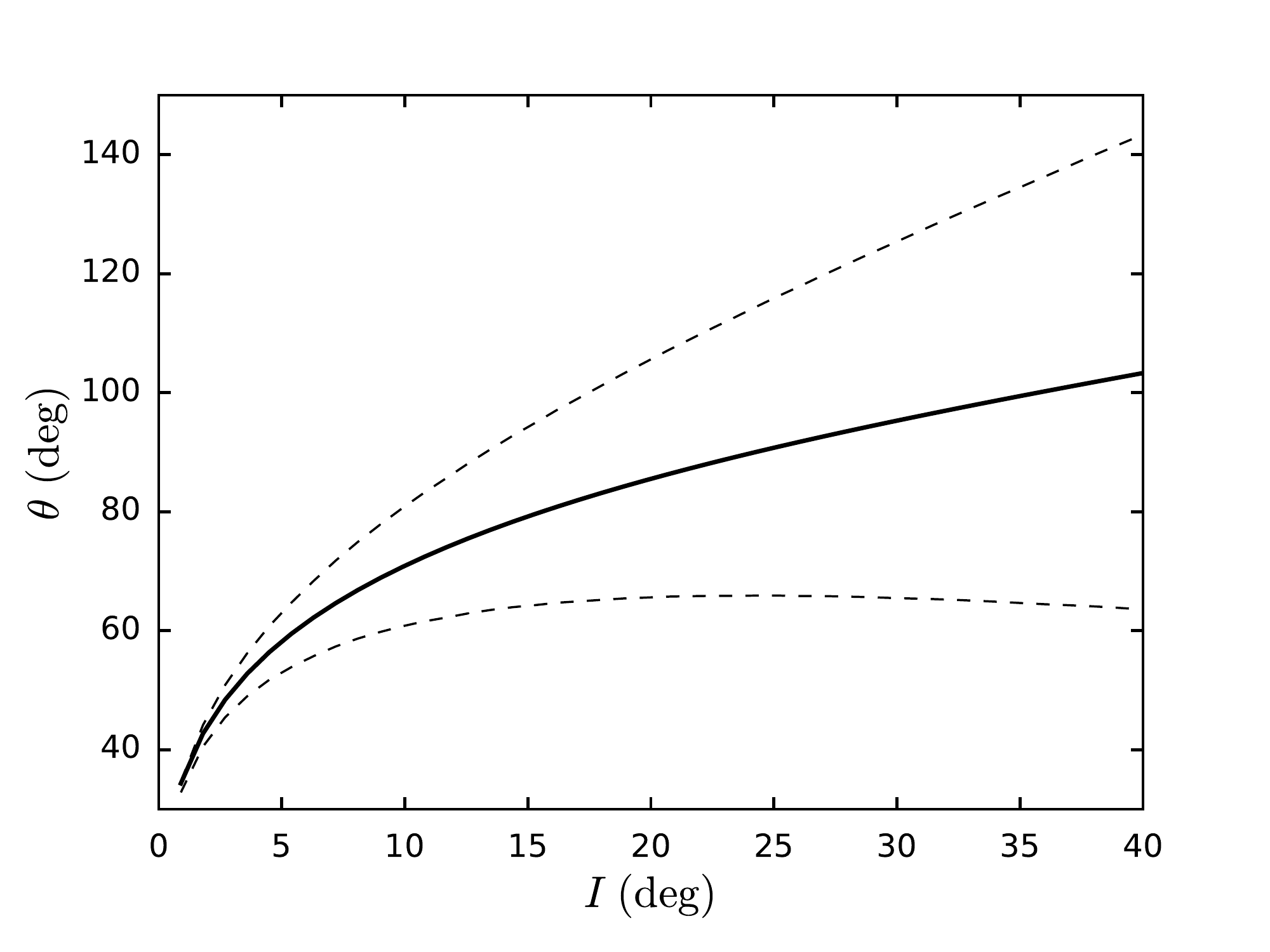}
\caption{Solid curve: Average value of $\theta$ following resonant
  excitation, as calculated from the area of the separatrix when
  $\theta_1$ and $\theta_4$ merge (equation [\ref{eq:theta_ave}]).  Dashed curves: variation of
  $\theta$ obtained from numerical integrations.  Since the system is
  in the weak-coupling regime after resonant excitation, the final obliquity variation
is simply $2 I$.}
\label{fig:theta_range}
\end{figure}

Since the phase-space area following the separatrix crossing is simply the area enclosed by the separatrix itself when $\eta = \eta_{\rm
  crit}$ (denoted as $A_{\rm sep}$), the final (when $\eta \gg 1$), average value of $\theta$ can be estimate from
\be
(\cos \theta)_{\rm ave} \simeq \frac{A_{\rm sep}}{2 \pi}.
\label{eq:theta_ave}
\ee
Since the spin and orbit are weakly coupled when $\eta \gg 1$, the range of obliquity variation (centered around $\theta_{\rm ave}$) is simply $2 I$.  Fig.~\ref{fig:theta_range} shows $\theta_{\rm ave}$ as a function of inclination, as determined by equation (\ref{eq:theta_ave}), along with the range of obliquity variation when $\eta \gg 1$, obtained from numerical integrations.  Equation (\ref{eq:theta_ave}) well captures the ``average'' value of obliquity following resonant excitation. 

In the example shown in Fig.~\ref{fig:time_ev_phase_space}, $\svec$ and $\lvec$ were initially aligned and librating around $\theta_1$.  When $\svec$ and $\lvec$ are initially slightly misaligned and circulating around $\theta_1$ (with small initial obliquity, $\theta_0 \lesssim 10^\circ$), the spin axis is eventually be captured into libration around $\theta_1$, after which the evolution proceeds very similarly to the case with zero initial obliquity.  Thus, the post-resonant obliquity variation does not depend sensitively on the initial obliquity, as long as the initial obliquity is not very large.

\section{Spin-Orbit Dynamics for Comparable ${\bf S}_\star$ and ${\bf L}_1$}
\label{sec:compSL}
The previous section considered the idealized case where $S_\star \ll L_1$, so that the torque from $\svec$ on $\lvec$ vanishes.  This simplified problem serves as a
useful reference point in understanding the dynamics of systems with comparable $S_\star$ and  $L_1$.  For the HJ/WJ systems of interest in
this paper $S_\star$ and $L_1$ may be comparable for rapidly rotating stars, although nearly always satisfying $S_\star \lesssim L_1$.  For the remainder of the paper we undertake numerical integrations of the ``real'' system, accounting for the torque on $\lvec$ due to $\svec$ (see equations [\ref{eq:dsdt}] - [\ref{eq:dlpdt}]), while allowing the stellar spin to decrease via magnetic braking according to equation (\ref{eq:magbraking}). In Section \ref{sec:gencassini} we extend the previous Cassini state analysis and derive results for generalized Cassini states, accounting for the effect of the spin on the orbit of the inner planet.  We show that this ``real'' system behaves qualitatively similar to the idealized problem, with a similar Cassini state transition coinciding with resonant obliquity growth.  In Section \ref{sec:numerical} we undertake numerical integrations and obtain quantitative results for generating spin-orbit misalignment for HJs and WJs with external companions of varying properties.

\subsection{Cassini States for Finite ${\bf S}_\star / {\bf L}_1$ and an Evolution Example}
\label{sec:gencassini}
\begin{figure}
\centering 
\includegraphics[scale=0.45]{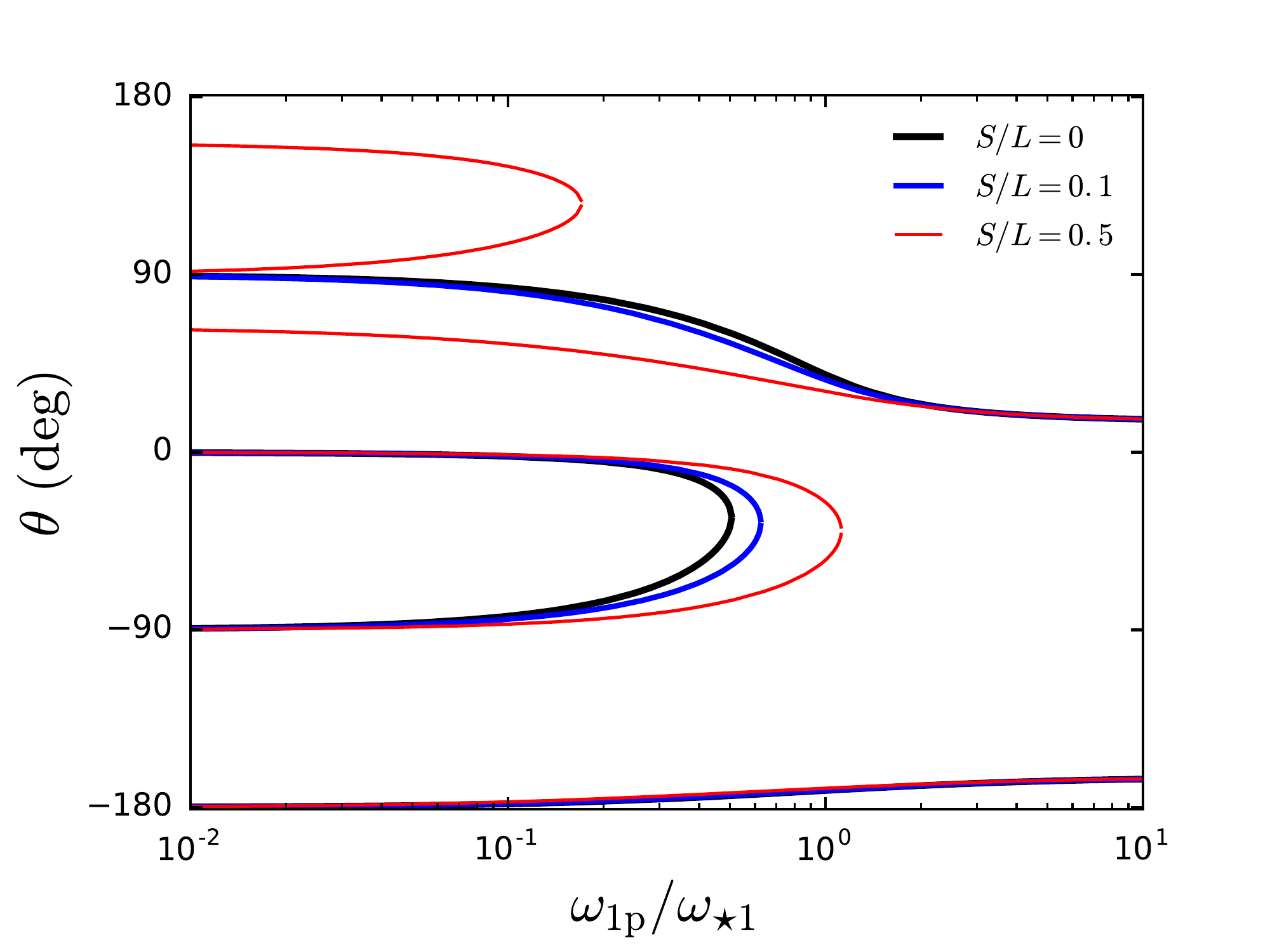}
\caption{Generalized Cassini state obliquities as a function of the coupling parameter $\omega_{1 {\rm p}} / \omega_{\star 1}$, obtained from equation (\ref{eq:cassini_gen}).  We have fixed $I = 20^\circ$ and $L_1 / L_{\rm p} = 0.3$.  The number of Cassini states as a function of coupling strength, as well as the obliquity values themselves are qualitatively similar for different values of $S_\star/L_1$.  For $S_\star/L_1 = 0.5$, additional retrograde equilibrium states exist for $\omega_{1 \rm p} / \omega_{\star 1} \lesssim 0.2$.}
\label{fig:gen_cass}
\end{figure}

Generalized Cassini states when $S_\star$ and $L_1$ are comparable were studied before by \cite{boue2006} and \cite{correia2015}.  In equilibrium, $\svec$, $\lvec$, and $\lpvec$
are coplanar, as in the case when $S_\star \ll L_1$.  This coplanar
configuration must be maintained through time ($\svec$, $\lvec$, and $\lpvec$
simply precess as a fixed plane in inertial space). We therefore require
\be
\frac{\D}{\D t} \bigg[ \svec \cdot
(\lvec \times \lpvec) \bigg] = 0.
\label{eq:theta}
\ee
After some algebra and substituting in the equations of motion (see equations [\ref{eq:dsdt}] - [\ref{eq:dlpdt}]), the equilibrium condition in equation (\ref{eq:theta}) can be written as
\ba
& & \frac{\omega_{1 {\rm p}}}{\omega_{\star 1}} \cos I \bigg[\cos I \cos(\theta - I) - \cos
\theta \bigg]  \nonumber \\
& & +  \frac{S_\star}{L_1} \cos \theta \bigg[\cos I - \cos(\theta -
I) \cos \theta \bigg] \nonumber \\
& & - \sin I \sin \theta \bigg[\cos \theta - \frac{L_1}{L_{\rm p}}
\frac{\omega_{1 {\rm p}}}{\omega_{\star 1}} \cos I \bigg] = 0.
\label{eq:cassini_gen}
\ea
Equation (\ref{eq:cassini_gen}) specifies the Cassini state obliquities, valid for general $S_\star / L_1$, and $L_1 / L_{\rm p}$.  In the limits 
$S_\star/L_1 \ll 1$ and $L_1/L_{\rm p} \ll 1$, equation (\ref{eq:cassini_gen}) reduces to equation (\ref{eq:cass}).  Fig.~\ref{fig:gen_cass} shows the generalized Cassini states as a function of $\omega_{1 {\rm p}} / \omega_{\star 1} \propto \epsilon_{\star 1}$, for a fixed $I = 20^\circ$, $L_1/L_{\rm p} = 0.3$, and various values of $S_\star/L_1$.   Fixing the ratio $S_\star/L_1$ while varying $\omega_{1 {\rm p}} / \omega_{\star 1}$ is admittedly somewhat artificial, but allows for a straightforward comparison with the case of $S_\star / L_1 = 0$ explored previously in Section 3.  The number of Cassini states as a function of coupling strength, as well as the obliquity values are qualitatively similar for different values of $S_\star/L_1$. When $S_\star/L_1$ is of order unity, additional retrograde equilibrium states emerge, but they are not expected to strongly affect the obliquity evolution for systems that start out with spin-orbit alignment, as considered in this paper.

\begin{figure}
\centering 
\includegraphics[scale=0.65]{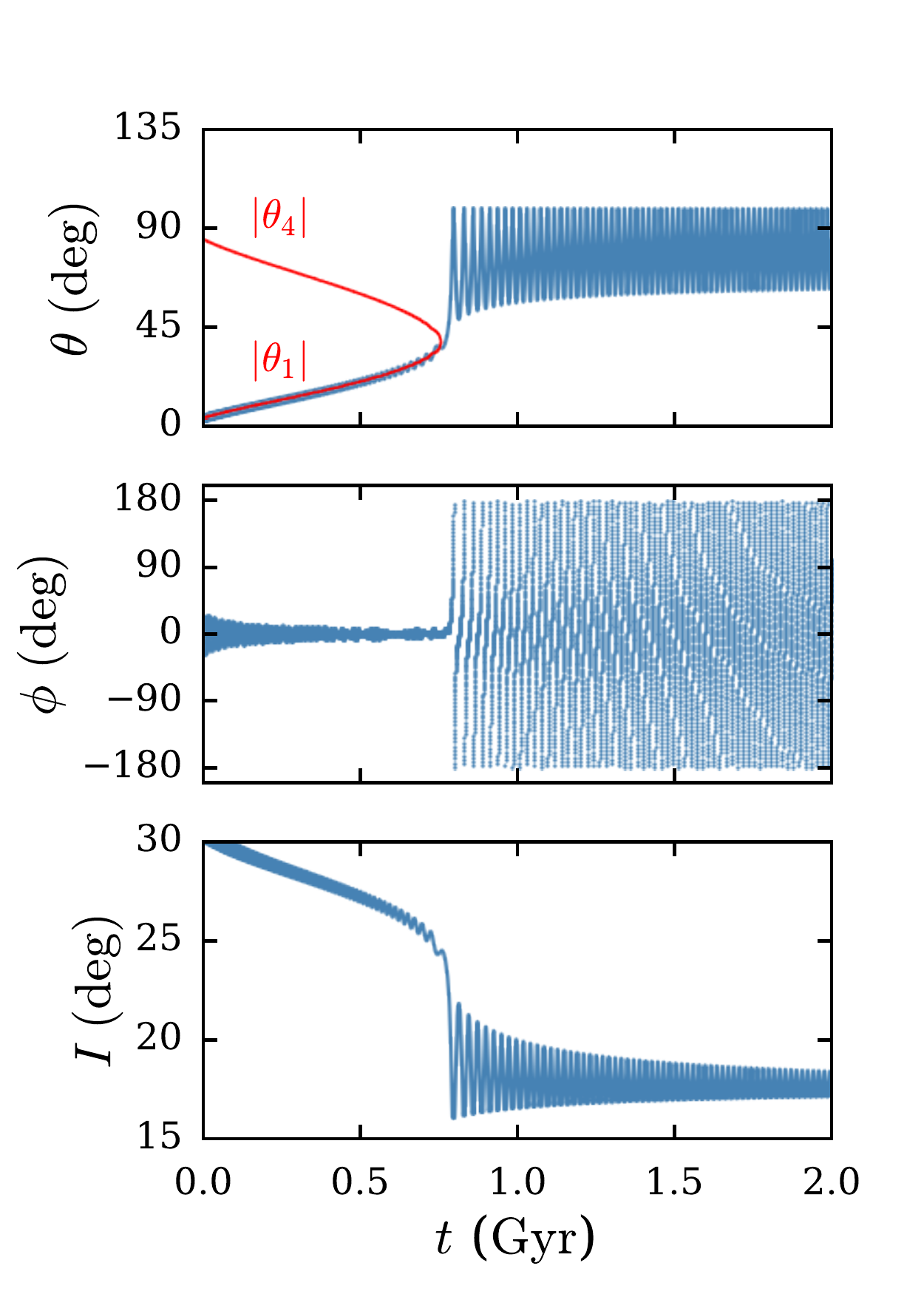}
\caption{Example of resonant obliquity excitation for a system with finite $S_\star$ and $L_1$.  The top panel shows the obliquity $\theta$, the middle panel shows the phase of $\svec$ relative to $\lvec$ ($\phi$), and the bottom panel shows the mutual inclination between $m_1$ and $m_p$ ($I$). The inner planet is a WJ, with $m_1 = M_J$, $a_1 = 0.2$ AU, and the external perturber has $\mper = M_J$, $\ap = 10$ AU, and $\ep = 0$, with an initial inclination (with respect to the orbit of $m_1$) $I_0 = 30^\circ$. The initial stellar spin period is $P_{\star,0} = 3$ days.  As the stellar spin decreases due to magnetic braking, $\theta$ evolves in a manner qualitatively similar to the idealized example shown in Fig.~\ref{fig:time_ev_phase_space}, with the spin axis librating around the instantaneous Cassini state 1 ($\theta_1$).  Eventually $\theta_1$ merges with $\theta_4$, and the obliquity is excited to a large value.  The increase in obliquity is accompanied by a decrease in mutual inclination.  }
\label{fig:example_feedback}
\end{figure}

Figure \ref{fig:example_feedback} shows an example of resonant obliquity excitation for an inner planet with $m_1 = M_J$, $a_1 = 0.2$ AU, and an external perturber with $\mper = M_{\rm J}$, $\ap = 10$ AU, and $\ep = 0$.  The mutual inclination between the two planets is initially $I_0 = 30^\circ$, and the stellar spin period is initially $P_{\star,0} = 3$ days, so that $\epsilon_{\star 1} \simeq 0.1$ at the start of the integration.  Inspection of Fig.~\ref{fig:example_feedback} reveals that the obliquity evolution is quite similar to the idealized example presented in Fig.~\ref{fig:time_ev_phase_space}: At early times, the spin and orbit are strongly coupled, and the spin axis librates closely around the instantaneous Cassini state 1 ($\theta_1$, as determined by equation [\ref{eq:cassini_gen}]).  Eventually, when the coupling parameter $\epsilon_{\star 1}$ becomes of order unity, the Cassini states $\theta_1$ and $\theta_4$ merge.   At this point, the obliquity jumps to a large value. Following this resonant excitation, when the spin and orbit become more weakly coupled, the obliquity oscillates between a minimum and maximum value.

One new feature in the dynamical evolution that emerges when $S_\star \sim L_1$ (and not captured in the idealized problem discussed in Section 3), is damping of the mutual inclination. As is evident from the bottom panel of Fig.~\ref{fig:example_feedback}, the mutual inclination decreases with time, with a sharp decline at $t \simeq 0.8$ Gyr, coinciding with the resonant obliquity growth.  This decrease in inclination can be understood as follows: The system initially librates around the Cassini state $\theta_1$, with $\theta_1 < 0$, so that $\svec$, $\lvec$, and $\lpvec$ are (in an average sense) coplanar, with $\svec$ and $\lpvec$ located on the opposite sides of $\lvec$; see Fig.~\ref{fig:vectors}.  As $\epsilon_{\star 1}$ increases (due to stellar spin-down), $|\theta_1|$ increases, so that $\svec$ and $\lvec$ are pushed apart.  This then implies that $\lvec$ is pushed closer to $\lpvec$, and $I$ must decrease.  By manipulating equations (\ref{eq:dsdt})-(\ref{eq:dlpdt}), we can derive expressions for $\D(\svec \cdot \lvec)/ \D t = \D \cos \theta / \D t$ and $\D(\lvec \cdot \lpvec)/ \D t = \D \cos I / \D t$, yielding the change in $I$ compared to the change in $\theta$:
\be
\frac{\D I}{\D \theta} = - \frac{S_\star}{L_1} \frac{\omega_{\star 1}}{\omega_{1 {\rm p}}} \left(\frac{\cos \theta \sin \theta}{\cos I \sin I} \right).
\ee
At late times, once the star has spun down, the quantity $S_\star \omega_{\star 1} / L_1 \omega_{1 {\rm p}}$ becomes small, and the decrease in inclination ceases, although the inclination may still undergo oscillations.  Resonant excitation of stellar obliquities thus tends to erase the mutual inclination between the inner planet and outer perturber.

\subsection{Results for HJs and WJs with External Companions}
\label{sec:numerical}
Having demonstrated in Fig.~\ref{fig:example_feedback} a typical example of resonant obliquity excitation (accompanied by a decrease in mutual orbital inclination), we next explore the parameter space for HJs/WJs with external companions of varying properties.  All results in this section have been initialized with aligned stellar spin and orbital axes ($\theta = 0$).   

To start, we set both the planet and perturber masses to $1 M_J$, and consider first an inner planet with $a_1 = 0.05$ AU (a canonical HJ) and next an inner planet with $a_1 = 0.2$ AU (a canonical WJ).  We set the initial stellar spin period to $P_{\star,0} = 3$ days, and explore various initial inclinations ($I_0 = 10^\circ - 40^\circ$) and perturber semi-major axes.  In all cases, we integrate the equations of motion for a timespan of $5$ Gyr\footnote{Since the spin-down rate is quite slow after $\sim 1$ Gyr due to the $P_\star \propto t^{1/2}$ dependence, these results are not particular sensitive to the chosen integration timespan of 5 Gyr.}, and record the ``final'' (between 4.5 - 5 Gyr) range of variation of the spin-orbit angle, $\min(\theta)$, $\max(\theta)$, and the final variation of the mutual orbital inclination, $\min(I)$, $\max(I)$.

\begin{figure*}
\centering 
\includegraphics[scale=0.95]{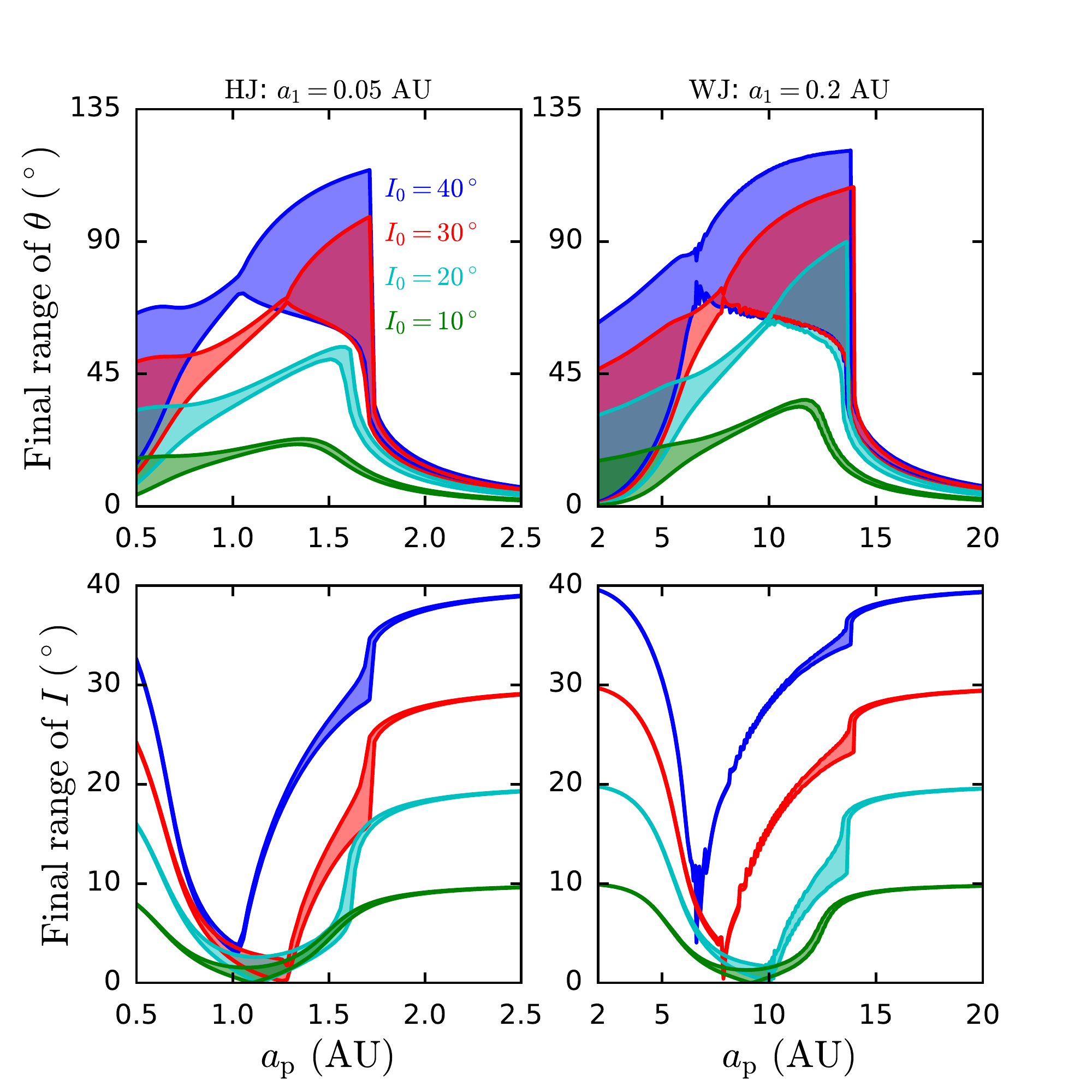}
\caption{Obliquity excitation as a function of perturber semi-major axis, showing various initial inclinations, as labeled.  The planet and perturber masses are $m_1 = \mper = 1 M_J$, and the initial stellar spin period is $P_{\star,0} = 3$ days.  The left panels show results for an inner planet with $a_1 = 0.05$ AU (a canonical HJ), and the right panels show an inner planet with $a_1 = 0.2$ AU (a canonical WJ).  The top panels show the final range of obliquities (between $\theta_{\rm min}$ and $\theta_{\rm max}$) at 4.5-5 Gyr, and the bottom panels show the final mutual inclination variation at 4.5-5 Gyr.  To excite substantial spin-orbit misalignment, an initial inclination $I_0 \gtrsim 20^\circ$ is needed, and the perturber must be located sufficiently close (see equation [\ref{eq:maxap}]).  Obliquity excitation (top panels) is accompanied by a decrease in mutual inclination (bottom panels).  In some instances, the initial inclination is almost completely erased.}
\label{fig:hwj}
\end{figure*}

Results for the canonical HJ case (with $a_1 = 0.05$ AU) are depicted in the left panels of Figure \ref{fig:hwj}.   For a close perturber with $\ap \lesssim 0.5$ AU, the spin and orbit are relatively weakly-coupled ($\epsilon_{\star 1} \gtrsim 1$) throughout the integration span.  After $5$ Gyr, the obliquity oscillates, with the final degree of variation depending on the initial mutual inclination, roughly in the range $0-2I$.  A more distant perturber, at $\ap \simeq 0.75 -1.75$ AU, induces resonant obliquity excitation, with the final variation of $\theta$ exhibiting a complicated dependence on $\ap$ and $I_0$. The obliquity excitation is often accompanied by a dramatic decrease in mutual inclination.  For example, when $I_0 = 30^\circ$ and $\ap \simeq 1 - 1.3$ AU, the final mutual inclination is less than $5^\circ$.   For perturbers beyond $\ap \simeq 1.75$ AU, the spin and orbit are always strongly coupled, so that the perturber is ineffective in exciting spin-orbit misalignment.  

The results for the canonical WJ case (with $a_1 = 0.2$ AU), shown in the right panels of Fig.~\ref{fig:hwj}, are qualitatively similar to those for the HJ case.  Given the larger value of $a_1$ for the WJ and the sensitive dependence of the spin precession on semi-major axis, resonant obliquity excitation may occur for much more distant (weaker) perturbers, with $\ap$ in the range $\sim 5 - 13$ AU.  For both the canonical HJ and WJ, a sufficient initial inclination is needed to generate a substantial obliquity.  For example, a perturber inclined by $10^\circ$ generates only a modest obliquity ($\lesssim 30^\circ$).  To produce a retrograde obliquity ($\gtrsim 90^\circ$) requires an initial inclination of at least $20^\circ - 30^\circ$.

Both the HJ and the WJ cases exhibit an abrupt decrease in obliquity excitation for perturbers beyond a maximum distance.  The maximum effective perturber semi-major axis $\tilde{a}_{\rm p, max}$ may be estimated by requiring that $\epsilon_{\star 1} (5 {\rm Gyr}) \gtrsim 1$, so that (see equation [\ref{eq:eps}])
\be
\tilde{a}_{\rm p, max} \simeq 1.5 \ {\rm AU} \bigg(\frac{a_1}{0.05 \ {\rm AU}} \bigg)^{3/2} \bigg(\frac{\mper}{m_1} \bigg)^{1/3}.
\label{eq:maxap}
\ee
For $\aptilde \gtrsim \tilde{a}_{\rm p, max}$, the perturber is unable to excite spin-orbit misalignment due to the strong coupling between $\svec$ and $\lvec$ throughout the stellar spin evolution.

Finally, we conduct a larger parameter survey, and examine the steady-state distribution of stellar obliquities and mutual orbital inclinations by plotting the values of $\theta$ and $I$ at a random time between $[0-5]$ Gyr.  We sample the parameters in the following ranges: $a_1 = [0.05 - 0.5]$ AU, $\ap = [10 - 100]a_1$, $e_{\rm p} = 0$, $\mper = [0.1 - 10] M_J$, $I_0= 10^\circ - 40^\circ$, and the initial stellar spin period in the observationally-motivated range $P_{\star,0} = 1 - 10$ days \citep[see, e.g. Fig.~1 of ][]{gallet2013}.  Given the large uncertainties in the statistical properties of long-period giant planets, this experiment is not meant to serve as a precise quantitative prediction for HJ/WJ obliquities, but rather to identify the orbital geometries that may lead to high obliquities.  We discard any systems that do not satisfy the stability condition given by \cite{petrovich2015}.  To avoid integrating systems that clearly will maintain spin-orbit alignment for the entire integration span, we also discard systems that satisfy $\epsilon_{\star 1} (P_\star = 30 {\rm d}) < 0.01$.

\begin{figure*}
\centering 
\includegraphics[width=\textwidth]{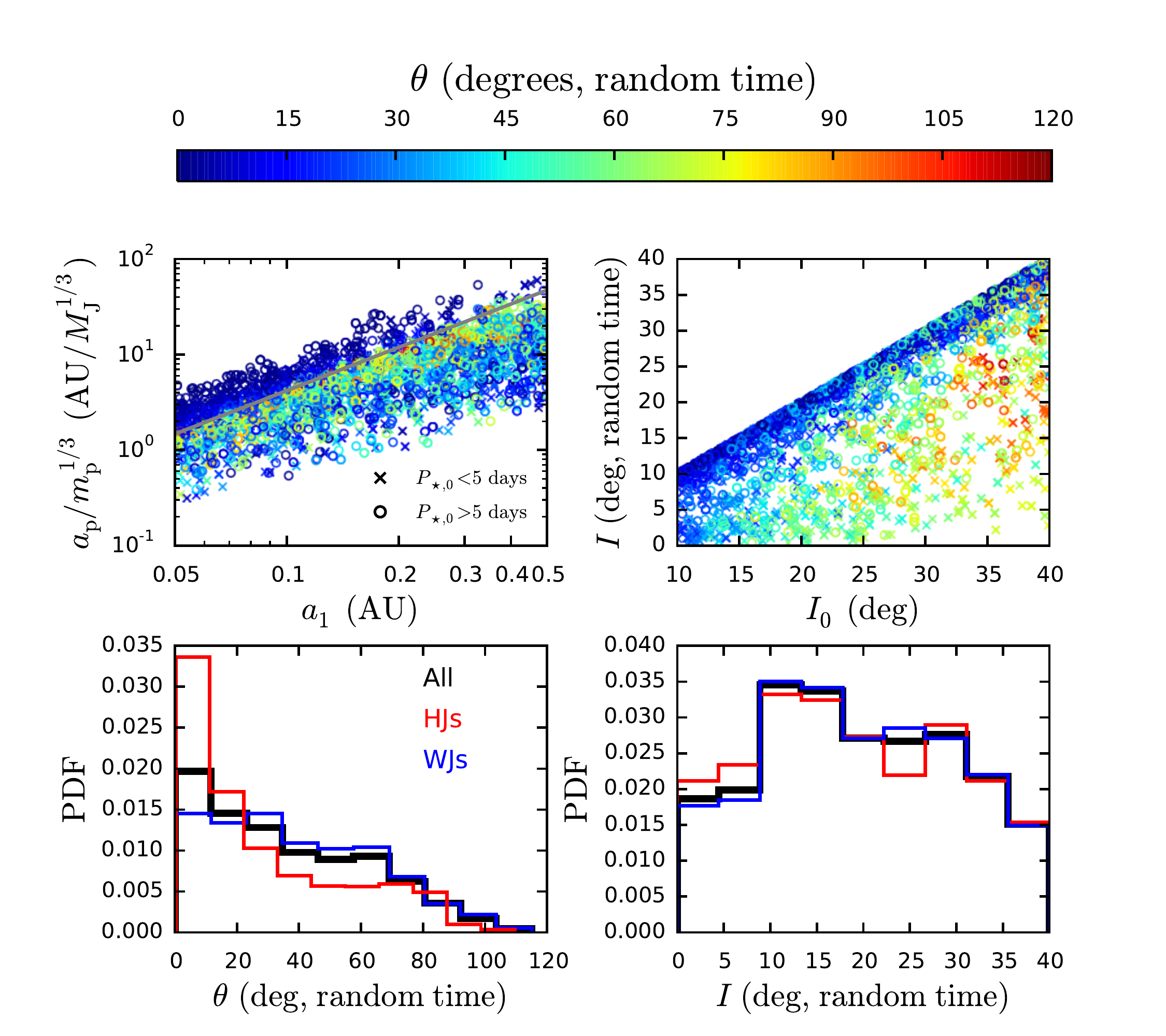}
\caption{Parameter survey of obliquity excitation and inclination decay in systems consisting of a host star, a HJ or WJ, and an external perturber.  We integrate the full secular equations of motion for a duration of time randomly chosen between $0 - 5$ Gyr and record the value of the obliquity $\theta$ (the angle between $\svec$ and $\lvec$) and inclination $I$ (the angle between $\lvec$ and $\lpvec$).  We fix the inner planet mass $m_1 = 1 M_J$ and vary the initial spin period uniformly in the range $P_{\star,0} = 1 -10$ days, the initial mutual inclination uniformly in $I_0 = 10^\circ - 40^\circ$, and the inner planetary semi-major axis in $a_1 = 0.05 - 0.5$ AU (where HJs are defined to have $a_1 < 0.1$ AU and WJs have $a_1 > 0.1$ AU).  We vary the perturber mass in the range $\mper = 0.1 - 10 M_J$ and the semi-major axis in the range $\ap = (10 - 100)a_1$.  {\it Top left:} Perturber ``strength'' $\ap/\mper^{1/3}$ vs inner planet semi-major axis $a_1$.  The color of the points indicates the obliquity $\theta$ at a random time.  The dashed grey line indicates the analytic estimate for the perturber strength in order to affect the obliquity (see equation [\ref{eq:maxap}]).  {\it Top right:} Steady-state inclination $I$ versus initial inclination $I_0$, illustrating how resonant obliquity excitation can erase mutual inclinations.  {\it Bottom panels:} Distributions of steady-state stellar obliquities (left), and mutual inclinations (right).  The thick black histograms show all systems, while the red (blue) histograms show results for HJs (WJs). }
\label{fig:random}
\end{figure*}

Figure \ref{fig:random} shows the results of this parameter survey.  The top left panel depicts the main result, with the perturber ``strength'' $\ap/\mper^{1/3}$ versus the inner planet semi-major axis $a_1$, and the color indicating the value of the obliquity at a random time.  The grey line shows the analytic estimate for the maximum perturber strength that may induce changes in the obliquity (see equation [\ref{eq:maxap}]).  The analytic estimate is in good agreement with the numerical results: Pertubers beyond $a_{\rm p,max}$ are unable to generate high obliquities.   The bottom left panel shows the steady-state distribution of obliquities (at a random time).   Recall that the initial distribution of obliquities is a $\delta$-function at $\theta = 0$.  Due to the presence of the perturber, a wide range of obliquities is generated, with a maximum obliquity of $\sim 113^\circ$.  For systems that undergo resonant excitation, the degree of obliquity excitation is highest for the weaker perturbers (near the grey line of the top left panel of Fig.~\ref{fig:random}).  This occurs because the amount of obliquity growth increases with decreasing $S_\star/L_1$.  Since the systems with weaker perturbers encounter the resonance at a later time (when $S_\star/L_1$ is smaller), such systems tend to result in higher obliquities.

The results for the steady state inclinations are depicted in the upper and lower right panels of Fig.~\ref{fig:random}.  Recall that the distribution of initial inclinations ($I_0$) is chosen to be uniform in $10^\circ - 40^\circ$.  Examining the distribution of inclinations (lower right panel), obliquity excitation leads to decreased inclinations, with a removal of points from the highest inclination bins ($\sim 30^\circ - 40^\circ$), and addition of points at the lowest inclination bins ($0 - 10^\circ$).  In some cases, the initial inclination is completely erased, as indicated in the upper right panel.

Recall that in this paper the range of initial inclinations is restricted to $I_0 < 40^\circ$, so that Lidov-Kozai eccentricity oscillations have no chance of developing.  However, the minimum inclination allowing Lidov-Kozai oscillations is often significantly larger than $40^\circ$, depending on the rate of apsidal precession due to GR compared to the apsidal precession due to the perturber \citep[e.g.][]{liu2015}.   As a result, qualitatively similar results to those shown in this paper may often occur for $I_0 > 40^\circ$, but with even larger excitement of obliquity.  Furthermore, if Lidov-Kozai cycles {\it do} arise, the evolution of the spin axis becomes chaotic \citep{storch2014, storch2015}, so that the full possible range of obliquities $(0^\circ - 180^\circ)$ may in some circumstances be reached.  

\section{Summary \& Discussion}
\label{sec:conclusion}
In this paper we have studied a new mechanism for generating
spin-orbit misalignments in HJ and WJ systems with external planetary
companions via resonant excitation.  Starting from initial
spin-orbit alignment, we evolve the stellar spin axis and the orbital
angular momentum axes of both planets (the HJ/WJ and outer perturber),
accounting for the gravitational torques between the inner and outer
planets and the oblate host star, as well as stellar spin-down due to magnetic
braking. For appropriate companion semi-major axis and mass (see
Fig.~\ref{fig:res_cond}), the inner system transitions from strong spin-orbit coupling
($\epsilon_{\star 1}\ll 1$; see equation [\ref{eq:eps}] for the definition of this ``coupling parameter'') at earlier times to weak coupling
($\epsilon_{\star 1} \gtrsim 1$) at later times as the star spins down. Appreciable 
stellar obliquity may be excited during resonance crossing
($\epsilon_{\star 1}\sim 1$),
when the stellar spin precession rate (around the inner planet) is 
comparable to the orbital precession rate of the inner planet 
(due to the outer perturber). 
Following the resonant obliquity growth, the spin and orbit are weakly coupled, and
the stellar obliquity oscillates between a minimum and a maximum,  whose
values depend on the details of the prior (pre-resonant) spin history.

Insight into the resonant growth of obliquity can be gained by
considering the idealized case where the stellar spin angular momentum
$S_\star$ is much less than the angular momentum of the inner planet
(HJ or WJ) $L_1$ (Section 3). In this case, the stellar spin axis
closely follows one of the Cassini (equilibrium) states, as stellar
spin-down gradually reduces the spin-orbit coupling strength,
until a separatrix crossing (at the resonance $\epsilon_{\star 1} \sim 1$)
leading to rapid obliquity excitation (see Fig.~\ref{fig:time_ev_phase_space}). The final average
value of spin-orbit misalignment can be computed analytically (see Fig.~\ref{fig:theta_range}).

For realistic HJ/WJ systems, $S_\star$ can be comparable to $L_1$, but
the spin-orbit dynamics remain qualitatively similar to the $S_\star\ll L_1$
case.  In particular, an initially aligned system follows a
generalized Cassini state (valid for arbitrary $S_\star/L_1$;  see Fig.~\ref{fig:gen_cass})
until the resonance crossing, leading to
rapid obliquity excitation (see Fig.~\ref{fig:example_feedback}).
An important new feature for systems with $S_\star\sim L_1$ is that 
the inclination angle between the inner planet and the outer companion 
tends to decrease as a result of obliquity growth (see Figs.~\ref{fig:example_feedback} - \ref{fig:hwj}).

Concerning spin-orbit misalignments of HJ and WJ systems, 
our main findings are as follows:

\begin{itemize}
\item Due to their close proximity to the host star, HJs have orbital axes that are strongly coupled to the host star spin axis (note the strong semi-major axis dependence in the coupling parameter $\epsilon_{\star 1}$ in equation [\ref{eq:eps}]).  As a result, for any kind of obliquity growth to be possible, a strong perturber is required (see equation [\ref{eq:maxap}] and Figs.~\ref{fig:res_cond} and \ref{fig:hwj} - \ref{fig:random}).  For example, a $1 M_{\rm J}$ HJ with semi-major axis $0.05$ AU requires that a $1 M_{\rm J}$ perturber be located within $\sim 1.75$ AU.

\item In contrast, the spin-orbit coupling in WJ systems is weaker, so that distant or low-mass perturbers may excite obliquities.  For example, a $1 M_{\rm J}$ WJ with semi-major axis $0.2$ AU requires that a $1 M_{\rm J}$ perturber be located within $\sim 13$ AU (see Figs.~\ref{fig:res_cond} and \ref{fig:hwj} - \ref{fig:random}).  

\item For both HJ and WJ systems, external perturbers must have modest inclinations ($I_0 \gtrsim 20^\circ$) in order to produce substantial obliquity growth (see Figs.~\ref{fig:hwj} and \ref{fig:random}).

\item Obliquity growth is generally accompanied by a decrease in mutual orbital inclination between the inner planet and outer perturber (see Figs.~\ref{fig:example_feedback} - \ref{fig:random}).  Resonant obliquity growth may thus erase high initial mutual inclinations in such systems.

\end{itemize}

This paper has focused on planetary companions to HJs/WJs, but stellar companions may also resonantly excite obliquities.  For HJs, a stellar-mass companion must be very close (within $\sim 10-20$ AU, due to the dependence on perturber properties as $\aptilde / \mper^{1/3}$).  Since such close stellar companions may inhibit planet formation in the first place \citep{wang2014}, it is unclear to what extent they contribute to HJ obliquities.  In contrast, more distant ($\sim$ hundred AU) stellar-mass companions to WJs may easily lead to resonant obliquity excitation.  Such stellar companions may be especially effective because they are expected to follow an isotropic distribution in inclination, so that a substantial fraction of binary perturbers may have high inclinations.

Throughout this paper, we have fixed the mass of the host star to $M_\star = 1 M_\odot$.  Hot Jupiter obliquities exhibit a well-known dependence on stellar effective temperature \citep{winn2010}, with HJs around cool stars ($T_{\rm eff} \lesssim 6200$ K) having low obliquities, and HJs around hot stars having high obliquities (see \citealt{winn2017} and \citealt{munoz2018} for recent discussions and statistics of this trend).  Hot stars do not experience strong magnetic braking (likely due to the absence of a surface convective zone), and remain rapidly rotating throughout their lifetimes.  As a result, resonant obliquity excitation is unlikely to occur around hot stars, because it requires that the perturber properties be somewhat fine-tuned.  Thus, the dependence of resonant excitation on stellar effective temperature appears to yield the opposite trend compared to observations.  This fact, together with the requirement that HJs need quite strong perturbers to have their obliquities raised at all, implies that resonant obliquity excitation is certainly not the entire story in HJ obliquities.  However, it may nonetheless be at work in individual misaligned systems.  Indeed, exceptions to the observed obliquity-effective temperature correlation do exist (e.g. WASP-8b, \citealt{queloz2010}, WASP-2b, \citealt{triaud2010}).

The story for WJs may be very different.  As noted previously in Section 1, a large fraction of WJs are observed to have external giant planet companions \citep{bryan2016}, many of which have the appropriate combinations of semi-major axis ($\sim 5 - 20$ AU) and mass to cause resonant obliquity excitation. Provided that such companions are sufficiently inclined, we predict that many WJs around cold stars have significant stellar spin-orbit misalignments due to resonant excitation, whereas hot stars would not have their obliquities resonantly excited, and tend to have low obliquities.  So far, WJ stellar obliquities are largely un-probed.  In the near future, NASA's TESS mission \citep{ricker2014} will discover a large number of WJs/HJs around bright stars. These systems will be amenable to Rossiter-McLaughlin measurements of spin-orbit misalignments, in addition to providing better statistics on the orbital parameters.  These new observations will help determine whether resonant obliquity excitation by external companions play an important role in WJ systems.

Regardless of the extent to which WJ obliquities are probed in the near future, resonant obliquity excitation has interesting implications for exoplanetary systems, due to the possibility that high initial inclinations can be erased.  The mutual inclinations ($\gtrsim 20^\circ$) needed for resonant excitation must be generated either via a scattering event of three or more giant planets or perturbations from a stellar companion.  As observations continue to constrain mutual inclinations in multi-planet systems, it is useful to keep in mind that such inclinations may not reflect the ``initial'' (i.e. previously higher inclinations following a scattering event or excitation from a nearby star), if resonant obliquity excitation has occurred.

\section*{Acknowledgments}
We thank Matija \'{C}uk, Doug Hamilton, and Dan Fabrycky for useful discussions.  This work has been supported in part by NASA grants NNX14AG94G and NNX14AP31G, and NSF grant AST-1715246.  K.R.A. is supported by the NSF Graduate Research Fellowship Program under Grant No. DGE-1650441.  

{}


\begin{thebibliography}{}

\bibitem[Albrecht et al.(2012)]{albrecht2012} 
Albrecht, S., Winn, J.~N., Johnson, J.~A., et al.\ 2012, ApJ, 757, 18

\bibitem[Anderson et al.(2016)]{anderson2016} 
Anderson, K.~R., Storch, N.~I., \& Lai, D.\ 2016, MNRAS, 456, 3671

\bibitem[Anderson et al.(2017)]{anderson2017} 
Anderson, K.~R., Lai, D., \& Storch, N.~I.\ 2017, MNRAS, 467, 3066

\bibitem[Anderson \& Lai(2017)]{AL2017} 
Anderson, K.~R., \& Lai, D.\ 2017, MNRAS, 472, 3692

\bibitem[Antonini et al.(2016)]{antonini2016} 
Antonini, F., Hamers, A.~S., \& Lithwick, Y.\ 2016, AJ, 152, 174

\bibitem[Barker \& Ogilvie(2009)]{barker2009} 
Barker, A.~J., \& Ogilvie, G.~I.\ 2009, MNRAS, 395, 2268

\bibitem[Bate et al.(2010)]{bate2010} 
Bate, M.~R., Lodato, G., \& Pringle, J.~E.\ 2010, MNRAS, 401, 1505

\bibitem[Batygin(2012)]{batygin2012} 
Batygin, K.\ 2012, Nature, 491, 418

\bibitem[Batygin \& Adams(2013)]{batygin2013} 
Batygin, K., \& Adams, F.~C.\ 2013, ApJ, 778, 169

\bibitem[Beaug{\'e} \& Nesvorn{\'y}(2012)]{beauge2012} 
Beaug{\'e}, C., \& Nesvorn{\'y}, D.\ 2012, ApJ, 751, 119

\bibitem[Becker et al.(2017)]{becker2017} 
Becker, J.~C., Vanderburg, A., Adams, F.~C., Khain, T., \& Bryan, M.\ 2017, AJ, 154, 230

\bibitem[Bolmont \& Mathis(2016)]{bolmont2016} 
Bolmont, E., \& Mathis, S.\ 2016, Celestial Mechanics and Dynamical Astronomy, 126, 275 

\bibitem[Bou{\'e} \& Laskar(2006)]{boue2006} 
Bou{\'e}, G., \& Laskar, J.\ 2006, Icarus, 185, 312

\bibitem[Bou{\'e} \& Fabrycky(2014)]{boue2014} 
Bou{\'e}, G., \& Fabrycky, D.~C.\ 2014, ApJ, 789, 111

\bibitem[Bouvier(2013)]{bouvier2013} 
Bouvier, J.\ 2013, EAS Publications Series, 62, 143

\bibitem[Bryan et al.(2016)]{bryan2016} 
Bryan, M.~L., Knutson, H.~A., Howard, A.~W., et al.\ 2016, ApJ, 821,
89

\bibitem[Colombo(1966)]{colombo1966} 
Colombo, G.\ 1966, AJ, 71, 891

\bibitem[Correia(2015)]{correia2015} 
Correia, A.~C.~M.\ 2015, A\&A, 582, A69

\bibitem[Dawson \& Johnson(2018)]{dawson2018} 
Dawson, R.~I., \& Johnson, J.~A.\ 2018, arXiv:1801.06117

\bibitem[Fabrycky \& Tremaine(2007)]{fabrycky2007b} 
Fabrycky, D., \& Tremaine, S.\ 2007, ApJ, 669, 1298 

\bibitem[Fabrycky et al.(2007)]{fabrycky2007} 
Fabrycky, D.~C., Johnson, E.~T., \& Goodman, J.\ 2007, ApJ, 665, 754

\bibitem[Fielding et al.(2015)]{fielding2015} 
Fielding, D.~B., McKee, C.~F., Socrates, A., Cunningham, A.~J., 
\& Klein, R.~I.\ 2015, MNRAS, 450, 3306

\bibitem[Foucart \& Lai(2011)]{foucart2011} 
Foucart, F., \& Lai, D.\ 2011, MNRAS, 412, 2799

\bibitem[Gallet \& Bouvier(2013)]{gallet2013} 
Gallet, F., \& Bouvier, J.\ 2013, A\&A, 556, A36

\bibitem[Hamers et al.(2017)]{hamers2017} 
Hamers, A.~S., Antonini, F., Lithwick, Y., Perets, H.~B., \& Portegies Zwart, S.~F.\ 2017, MNRAS, 464, 688 

\bibitem[H{\'e}brard et al.(2008)]{hebrard2008} 
H{\'e}brard, G., Bouchy, F., Pont, F., et al.\ 2008, A\&A, 488, 763

\bibitem[Heller(2018)]{heller2018} 
Heller, R.\ 2018, arXiv:1806.06601

\bibitem[Henrard \& Murigande(1987)]{henrard1987} 
Henrard, J., \& Murigande, C.\ 1987, Celestial Mechanics, 40, 345

\bibitem[Huang et al.(2016)]{huang2016} 
Huang, C., Wu, Y., \& Triaud, A.~H.~M.~J.\ 2016, ApJ, 825, 98

\bibitem[Kaib et al.(2011)]{kaib2011} 
Kaib, N.~A., Raymond, S.~N., \& Duncan, M.~J.\ 2011, ApJL, 742, L24

\bibitem[Lai et al.(2011)]{lai2011} 
Lai, D., Foucart, F., \& Lin, D.~N.~C.\ 2011, MNRAS, 412, 2790

\bibitem[Lai(2012)]{lai2012} 
Lai, D.\ 2012, MNRAS, 423, 486

\bibitem[Lai(2014)]{lai2014} 
Lai, D.\ 2014, MNRAS, 440, 3532

\bibitem[Lai \& Pu(2017)]{lai2017} 
Lai, D., \& Pu, B.\ 2017, AJ, 153, 42

\bibitem[Lai, Anderson, \& Pu(2018)]{lai2018} 
Lai, D., Anderson, K.~R., \& Pu, B.\ 2018, MNRAS, 475, 5231

\bibitem[Liu et al.(2015)]{liu2015} 
Liu, B., Mu{\~n}oz, D.~J., \& Lai, D.\ 2015, MNRAS, 447, 747

\bibitem[Mardling(2010)]{mardling2010} Mardling, R.~A.\ 2010, MNRAS, 407, 1048

\bibitem[Masuda(2017)]{masuda2017} 
Masuda, K.\ 2017, AJ, 154, 64

\bibitem[Mathis(2015)]{mathis2015} 
Mathis, S.\ 2015, A\&A, 580, L3 

\bibitem[Mills \& Fabrycky(2017)]{mills2017} 
Mills, S.~M., \& Fabrycky, D.~C.\ 2017, AJ, 153, 45 

\bibitem[Mu{\~n}oz et al.(2016)]{munoz2016} 
Mu{\~n}oz, D.~J., Lai, D., \& Liu, B.\ 2016, MNRAS, 460, 1086

\bibitem[Mu{\~n}oz \& Perets(2018)]{munoz2018} 
Mu{\~n}oz, D.~J., \& Perets, H.~B.\ 2018, arXiv:1805.03654

\bibitem[Nagasawa et al.(2008)]{nagasawa2008} 
Nagasawa, M., Ida, S., \& Bessho, T.\ 2008, ApJ, 678, 498

\bibitem[Naoz et al.(2012)]{naoz2012} Naoz, S., Farr, W.~M., \& Rasio, F.~A.\ 2012, ApJL, 754, L36

\bibitem[Narita et al.(2009)]{narita2009} 
Narita, N., Sato, B., Hirano, T., \& Tamura, M.\ 2009, PASJ, 61, L35

\bibitem[Ogilvie(2013)]{ogilvie2013} 
Ogilvie, G.~I.\ 2013, MNRAS, 429, 613 

\bibitem[Peale(1969)]{peale1969} 
Peale, S.~J.\ 1969, AJ, 74, 483 

\bibitem[Peale(1974)]{peale1974} 
Peale, S.~J.\ 1974, AJ, 79, 722

\bibitem[Perryman et al.(2014)]{perryman2014} 
Perryman, M., Hartman, J., Bakos, G.~{\'A}., \& Lindegren, L.\ 2014, ApJ, 797, 14

\bibitem[Petrovich(2015a)]{petrovich2015a} 
Petrovich, C.\ 2015a, ApJ, 805, 75  

\bibitem[Petrovich(2015b)]{petrovich2015b} 
Petrovich, C.\ 2015b, ApJ, 799, 27  

\bibitem[Petrovich(2015c)]{petrovich2015} 
Petrovich, C.\ 2015c, ApJ, 808, 120

\bibitem[Pu \& Lai(2018)]{pu2018} 
Pu, B., \& Lai, D.\ 2018, MNRAS, 478, 197 

\bibitem[Queloz et al.(2010)]{queloz2010} 
Queloz, D., Anderson, D.~R., Collier Cameron, A., et al.\ 2010, A\&A, 517, L1

\bibitem[Rasio \& Ford(1996)]{rasio1996} 
Rasio, F.~A., \& Ford, E.~B.\ 1996, Science, 274, 954

\bibitem[Ricker et al.(2014)]{ricker2014} Ricker, G.~R., Winn, J.~N., Vanderspek, R., et al.\ 2014, Space Telescopes and Instrumentation: Optical, Infrared, and Millimeter Wave, 9143, 914320

\bibitem[Skumanich(1972)]{skumanich1972} 
Skumanich, A.\ 1972, ApJ, 171, 565

\bibitem[Spalding \& Batygin(2014)]{spalding2014} 
Spalding, C., \& Batygin, K.\ 2014, ApJ, 790, 42

\bibitem[Storch et al.(2014)]{storch2014} 
Storch, N.~I., Anderson, K.~R., \& Lai, D.\ 2014, Science, 345, 1317

\bibitem[Storch \& Lai(2015)]{storch2015} 
Storch, N.~I., \& Lai, D.\ 2015, MNRAS, 448, 1821

\bibitem[Storch et al.(2017)]{storch2017} 
Storch, N.~I., Lai, D., \& Anderson, K.~R.\ 2017, MNRAS, 465, 3927

\bibitem[Triaud et al.(2010)]{triaud2010} 
Triaud, A.~H.~M.~J., Collier Cameron, A., Queloz, D., et al.\ 2010, A\&A, 524, A25

\bibitem[Wang et al.(2014)]{wang2014} 
Wang, J., Xie, J.-W., Barclay, T., \& Fischer, D.~A.\ 2014, ApJ, 783, 4

\bibitem[Ward et al.(1979)]{ward1979} 
Ward, W.~R., Burns, J.~A., \& Toon, O.~B.\ 1979, Geophys. Res., 84,
243 

\bibitem[Ward \& Hamilton(2004)]{ward2004} 
Ward, W.~R., \& Hamilton, D.~P.\ 2004, AJ, 128, 2501

\bibitem[Winn et al.(2005)]{winn2005} 
Winn, J.~N., Noyes, R.~W., Holman, M.~J., et al.\ 2005, ApJ, 631, 1215

\bibitem[Winn et al.(2009)]{winn2009} 
Winn, J.~N., Johnson, J.~A., Albrecht, S., et al.\ 2009, ApJL, 703, L99

\bibitem[Winn et al.(2010)]{winn2010} 
Winn, J.~N., Fabrycky, D., Albrecht, S., \& Johnson, J.~A.\ 2010, ApJL, 718, L145

\bibitem[Winn \& Fabrycky(2015)]{winn2015} 
Winn, J.~N., \& Fabrycky, D.~C.\ 2015, ARAA, 53, 409

\bibitem[Winn et al.(2017)]{winn2017} 
Winn, J.~N., Petigura, E.~A., Morton, T.~D., et al.\ 2017, AJ, 154, 270 

\bibitem[Wu \& Murray(2003)]{wu2003} 
Wu, Y., \& Murray, N.\ 2003, ApJ, 589, 605

\bibitem[Wu \& Lithwick(2011)]{wu2011} 
Wu, Y., \& Lithwick, Y.\ 2011, ApJ, 735, 109

\bibitem[Zahn(1977)]{zahn1977} Zahn, J.-P.\ 1977, A\&A, 57, 383

\bibitem[Zanazzi \& Lai(2018)]{zanazzi2017} 
Zanazzi, J.~J., \& Lai, D.\ 2018, MNRAS, 478, 835

\end{thebibliography}
\end{document}